\documentclass[showpacs,showkeys,preprintnumbers,nofootinbib,superscriptaddress,
amsmath,floatfix,amssymb,secnumarabic,aps,prd,a4paper]{revtex4}

\usepackage[colorlinks=true,breaklinks=false,urlcolor=blue]{hyperref}
\usepackage{graphicx}
\usepackage{epsfig}
\usepackage{multirow}

\newcommand{\beq}{\begin{equation}}
\newcommand{\eeq}{\end{equation}}
\newcommand{\beqar}{\begin{eqnarray}}
\newcommand{\eeqar}{\end{eqnarray}}

\def\dalam{\hbox
{\vrule\vbox{\hrule\hbox to 1ex{ \hfill}\kern 1 ex\hrule}\vrule}}

\def\1/2{\hbox{$ {1 \over 2}$ }}

\def\h{\hbar}
\def\i/h{{i \over \h}}

\def\ch{\cosh}
\def\sh{\sinh}

\def\inf{\infty}
\def\pd{\partial} 
\def\v{\vec}

\def\a{\alpha} 
\def\b{\beta}

\def\d{\delta} \def\D{\Delta}
\def\l{\lambda} 
\def\e{\epsilon}

\def\s{\sigma}

\def\c{\chi} 
\def\vf{\varphi}
 \def\F{\Phi}
 
\def\p{\psi}

\def\m{\mu}
\def\n{\nu}

\def\k{\kappa}

\def\W{\Omega}
\def\tt{\theta}

\def\<{\langle}
\def\>{\rangle}

\def\({\left(}
\def\[{\left[}
\def\){\right)}
\def\]{\right]}

\usepackage{subfigure}
\newcommand{\hm}[1]{#1\nobreak\discretionary{}{\hbox{\ensuremath{#1}}}{}}
\usepackage{tikz}
\usetikzlibrary{decorations.pathreplacing}
\newcommand{\myfrac}[2]{{\ifmmode{}^{#1}\!/_{\!#2}\else${}^{#1}\!/_{\!#2}$\fi}}

\usetikzlibrary{decorations.pathmorphing}    

\usepackage{dcolumn}		
\newcolumntype{.}{D{.}{.}{-1}}
\newcolumntype{b}[1]{D{.}{.}{#1}}
\newcommand{\mc}[1]{\multicolumn{1}{c}{#1}} 

\begin{document}
\sloppy

\title{Perturbativity vs non-perturbativity  in QED-effects for H-like atoms with  $Z\a >1$ }

\author{A.~Roenko}
\email{roenko@physics.msu.ru} \affiliation{Department of Physics and
Institute of Theoretical Problems of MicroWorld, Moscow State
University, 119991, Leninsky Gory, Moscow, Russia}

\author{K.~Sveshnikov}
\email{costa@bog.msu.ru} \affiliation{Department of Physics and
Institute of Theoretical Problems of MicroWorld, Moscow State
University, 119991, Leninsky Gory, Moscow, Russia}

\date{\today}

\begin{abstract}

The behavior of  levels near the threshold of the lower continuum in superheavy H-like atoms with $Z\a >1$, caused  by the interaction $\D U_{AMM}$ of the  electron's magnetic anomaly (AMM) dynamically screened at small distances $ \ll 1/m$,  with the  Coulomb field of atomic nucleus is considered by taking into account the complete dependence of electron's wavefunction (WF) on $Z\a$. It is shown that the calculation of the contribution caused by $\D U_{AMM}$ via both the  quark structure and the whole nucleus, considered as a uniformly charged extended Coulomb source, leads to  results, which  coincide within the accepted precision of calculations. It is shown also that there appears some difference in results between perturbative and non-perturbative methods of accounting for the contribution from $\D U_{AMM}$ within the corresponding Dirac equation (DE) in favor of the latter. Moreover, the growth rate of the contribution from $\D U_{AMM}$ reaches its maximum at $ Z \sim 140-150$, while by  further increase of $Z$ into the supercritical region $Z\gg Z_{cr,1}$ the shift of levels caused by $\D U_{AMM}$ near  the lower continuum decreases monotonically to zero. The last result is generalized to the whole self-energy contribution to the  shift of levels and so to the possible behavior of radiative QED-effects with virtual photon exchange  near  the lower continuum.

\end{abstract}


\maketitle

\section{Introduction\label{sec:1}}

The interaction of the Dirac fermion with the Coulomb field of a point-like source by taking into  account the fermionic AMM has been intensively explored in~\cite[and refs. therein]{Barut1975, Geiger1988, Barut1990, Reitz2000} in order to study the possibility of resonances in the systems like $e^+e^-$ at hadronic scales due to (possibly) increasing role of magnetic effects at extremely small distances. In the present paper we consider another aspect of the fermionic AMM, namely, the corresponding shifts of the lowest electronic levels in the field of an extended Coulomb source with extremely large $Z\a >1$ (like  superheavy nuclei) with non-perturbative account for the source charge and size. The keypoint here is that for  $Z > Z_{cr,1} \sim 170$ QED predicts a non-perturbative vacuum reconstruction, which should be followed by a series of nontrivial effects, including,  in particular, the vacuum positron emission~\cite[and refs. therein]{Greiner1985a, Ruffini2010,  Rafelski2016}. However, the long-term experiments at  GSI (Darmstadt) and Argonne National Lab didn't succeed in the  unambiguous conclusion of the status of the overcritical region, what promotes the question of the possible role of nonlinearity in QED effects for $Z > Z_{cr,1}$ to be quite actual~\cite{Rafelski2016, Ruffini2010, Schwerdtfeger2015, Sveshnikov2013, Davydov2017, Voronina2017}. In particular, the recent essentially non-perturbative results for the vacuum polarization energy for $Z > Z_{cr,1}$ show, that in the supercritical region the QED-effects could be substantially different from the perturbative case~\cite{Davydov2017, Voronina2017}. And although for an atomic electron $\D U_{AMM}$ is just a component of the self-energy contribution to the total radiative shift of the levels, it nevertheless occupies a special position, since it is described by a local operator, which preserves all the required for the Furry picture properties of the hamiltonian. So it allows for a detailed non-perturbative analysis, both in  $Z \a$ and (partially) in $\a/\pi$, since the latter enters as a factor in the coupling constant for $\D U_{AMM}$, and also the comparison  of results with those coming from PT.

It is well known, that the electronic AMM is a specific radiative effect, rather than  an immanent  property of the electron, hence, for strong external fields or  extremely small distances $\ll 1/m$ the dependence of the electronic formfactor $F_2(q^2)$ on the momentum transfer should be taken into account from the very beginning~\cite{Lautrup1976, Barut1977, Geiger1988}. At the same time, the effective Dirac-Pauli potential
\beq \label{eq:1}
\D U_{AMM}= \frac{\Delta g_{free}}{2} \frac{e}{ 4m}\, \s^{\m\n}F_{\m\n}\,  
\eeq
turns out to be correct  in the limit of extremely low momentum transfer only, when the former dependence could be ignored, i.e. $F_2(q^2)\simeq F_2(0)$. In the general case the calculation of the  formfactors responsible for AMM should be implemented via  self-consistent treatement of both    the external field  and  electronic WF~\cite{ Barut1977}. However, for the stationary electronic states even in superheavy atoms with  $Z > Z_{cr,1}$ the mean radius of the electronic WF substantially exceeds the nucleus size. So the time of the electron inside the nucleus doesn't exceed certain percents,  hence, the correct estimate for the corresponding  formfactors could be made within  PT in $\a/\pi$.   Since the one-loop correction to the vertex function   can be represented via electronic formfactors $F_1(q^2)$ and $F_2(q^2)$ in the form~\cite{Itzykson1980}
\begin{equation}\label{eq:2}
\Gamma^\mu (q^2) = \gamma^\mu F_1(q^2) + \frac{i}{2m} F_2(q^2) \sigma^{\mu\nu} q_\nu \, ,
\end{equation}
for strong  fields or  extremely small distances $\ll 1/m$  the effective potential~\eqref{eq:1} should be replaced by  expression
\begin{equation}\label{eq:3}
\D U_{AMM}(\vec{r}\,)= \frac{e}{ 2m}\, \s^{\m\n}\partial_\mu \mathcal{A}^{(cl)}_\nu(\vec{r}\,) ,
\end{equation}
where
\begin{equation}\label{eq:4}
\mathcal{A}^{(cl)}_\mu(\vec{r}\,) = \frac{1}{(2\pi)^3} \int \! d\vec{q} \ e^{i \vec{q}\,\vec{r}}\, \tilde{A}_\mu^{(cl)}(\vec{q}\,) F_2(-\vec{q}\,^2) \, ,
\end{equation}
while $\tilde{A}_\mu^{(cl)}(\vec{q}\,)$ is the Fourier-transform of the external field $A^{(cl)}_\mu(\vec{r}\,)$.
Accounting for the dependence on the momentum transfer leads to the behavior of the operator~\eqref{eq:3} for the point-like source  as $\sim\log m r$ for $r \to 0$~\cite{Lautrup1976}, while the Dirac-Pauli operator~\eqref{eq:1} reveals the maximally permissible for  DE  singularity $\sim 1/r^2$. In the last case the  particle WF becomes regular everywhere for any  $Z$,  acquiring  zeros of infinite  multiplicity for both components of bispinor in the Coulomb singularity, which substantially alter the results for the energy shifts~\cite{Sveshnikov2013}.

 Furthermore,  the genuine sources of electric charge in the extended nuclei should be the valence  (constituent)  $u$- and $d$-quarks, which reveal as the carriers of (fractional) charge the same properties as the atomic electrons, i.e. their charges should be always localized at  certain spatial points inside the nucleus. So  the consistent analysis of the contribution from $\D U_{AMM}$ in the superheavy nuclei requires for setting the nucleus structure as a discrete system of point-like (fractional) charges, spread over its volume. It is shown in this paper, how nontrivial such an analysis turns out to be, and how its results change by transition from the single point-like sources (quarks), placed in the nucleus center, to accounting for the contribution from the nucleus periphery, and also by transition from the purely perturbative approach to $\D U_{AMM}$ to the non-perturbative one. More concretely, it will be shown  that the dynamical screening of  AMM takes place first of all at small distances $\ll 1/m$, rather than due to the large magnitude of the external field.  As a consequence, calculation of the contribution from $\D U_{AMM}$ via both  the quark structure and the whole nucleus, considered as a uniformly charged extended Coulomb source, leads to  results coinciding within the accepted precision of calculations. At the same time, there appears a small, but remarkable difference in results between perturbative and non-perturbative methods of accounting for the contribution from $\D U_{AMM}$ within DE in favor of the latter. Moreover, the growth rate of the contribution from $\D U_{AMM}$ reaches its maximum at $ Z \sim 140-150$, while by  further increase of $Z$ into the supercritical region $Z\gg Z_{cr,1}$ the shift,  caused by $\D U_{AMM}$ for levels approaching the threshold of the lower continuum, decreases monotonically to zero.

\section{The effective interaction caused by AMM\label{sec:2}}

The effective interaction due to AMM~\eqref{eq:3} should be found in the next way. For an atomic electron the external potential takes the form
 $A^{(cl)}_\mu(\vec{r}\,)\hm=\delta_{0,\mu} \Phi(r)$, with $\Phi(r)$ being the spherically-symmetric Coulomb potential of the nucleus.
    Upon taking account that to the leading order $F_2(0) = {\a}/{2\pi} \simeq {\Delta g_{free}}/{2}$, one obtains after angular integration  in~\eqref{eq:4}
\begin{equation}\label{eq:6}
\mathcal{A}^{(cl)}_\mu(r) = -\frac{\Delta g_{free}}{2}\,\frac{Z e}{4\pi r} \, c(r)\, \delta_{\mu,0}, \quad c(r)=2 \int\limits_0^\infty \! q dq \ \sin q r \(-\frac{1}{Z e} \,\tilde{\Phi}(q) \) \frac{1}{\pi} \frac{F_2(-q^2)}{F_2(0)} \, .
\end{equation}
Now the effective potential~\eqref{eq:3} can be rewritten as
\begin{equation}\label{eq:7}
\D U_{AMM}(r)=-i\,Z\lambda\,\vec{\gamma}\cdot\vec{\nabla}\(-\frac{c(r)}{r}\)\, , 
\end{equation}
where $\lambda=\alpha^2/4\pi m$, $\a = e^2/4\pi$. In the next step, let us consider the calculation of the function  $c(r)$ for the given  $\Phi(r)$.

\subsection{The point-like source\label{sec:2a}}

\begin{figure}
\center
\subfigure[]{\label{pic:1a}
\begin{tikzpicture}
\draw[thin] (-2.2,0) -- (2.2,0);
\draw[thin] (0,-1) -- (0,1);
\draw[thin,decorate,decoration={snake,amplitude=1,segment length=5}] (0,1) -- (0,2.2);
\draw[thin,decorate,decoration={snake,amplitude=1,segment length=5}] (0,-1) -- (0,-2.2);
\draw[thick,rounded corners=3pt] (0,0) -- (2,0) arc (0:87:2) -- (0.1047,1);
\draw[thick] (0.1047,1) arc (0:-180:0.1047);
\draw[thick,<-,rounded corners=3pt] (0,0) -- (-2,0) arc (180:93:2) -- (-0.1047,1);
\filldraw[black] (0,.3) circle (1pt)
				(0,-.3) circle (1pt);
\draw (.1,0.3) node[right] {\small $i\mu$};
\draw (.1,-0.3) node[right] {\small -$i\mu$};
\draw (.1,1) node[right] {\small $2mi$};
\draw (.1,-1) node[right] {\small -$2mi$};
\draw (-1.8,1.8) node[draw] {\small $q$};
\end{tikzpicture}
}
\hspace{1em}
\subfigure[]{\label{pic:1b}
\begin{tikzpicture}
\draw[thin] (-2.2,0) -- (2.2,0);
\draw[thin] (0,-1) -- (0,1);
\draw[thin,decorate,decoration={snake,amplitude=1,segment length=5}] (0,1) -- (0,2.2);
\draw[thin,decorate,decoration={snake,amplitude=1,segment length=5}] (0,-1) -- (0,-2.2);
\draw[thick,rounded corners=3pt] (0,0) -- (2,0) arc (0:87:2) -- (0.1047,1);
\draw[thick] (0.1047,1) arc (0:-180:0.1047);
\draw[thick,<-,rounded corners=3pt] (0,0) -- (-2,0) arc (180:93:2) -- (-0.1047,1);
\filldraw[black] 
				(0,-.3) circle (1pt);
\draw (.1,-0.3) node[right] {\small -$i\mu$};
\draw (.1,1) node[right] {\small $2mi$};
\draw (.1,-1) node[right] {\small -$2mi$};
\draw (-1.8,1.8) node[draw] {\small $q$};
\end{tikzpicture}
}
\hspace{1em}
\subfigure[]{\label{pic:1c}
\begin{tikzpicture}
\draw[thin] (-2.2,0) -- (2.2,0);
\draw[thin] (0,-1) -- (0,1);
\draw[thin,decorate,decoration={snake,amplitude=1,segment length=5}] (0,1) -- (0,2.2);
\draw[thin,decorate,decoration={snake,amplitude=1,segment length=5}] (0,-1) -- (0,-2.2);
\draw[thick,rounded corners=3pt] (0,0) -- (2,0) arc (0:87:2) -- (0.1047,1);
\draw[thick] (0.1047,1) arc (0:-180:0.1047);
\draw[thick,<-,rounded corners=3pt] (0,0) -- (-2,0) arc (180:93:2) -- (-0.1047,1);
\draw[thick,rounded corners=3pt] (0,0) -- (2,0) arc (0:-87:2) -- (0.1047,-1);
\draw[thick] (0.1047,-1) arc (-180:0:-0.1047);
\draw[thick,<-,rounded corners=3pt] (0,0) -- (-2,0) arc (-180:-93:2) -- (-0.1047,-1);

\filldraw[black] (0,.3) circle (1pt);
\draw (.1,0.3) node[right] {\small $i\mu$};
\draw (.1,1) node[right] {\small $2mi$};
\draw (.1,-1) node[right] {\small -$2mi$};
\draw (-1.8,1.8) node[draw] {\small $q$};
\draw (.9,1.9) node[right] {\small $r<R$};
\draw (.9,-1.9) node[right] {\small $r>R$};
\end{tikzpicture}
}
\caption{The integration contours, used by calculation of the integrals
 in the formulae~\eqref{eq:9},~\eqref{eq:14}.}\label{pic:1}
\end{figure}
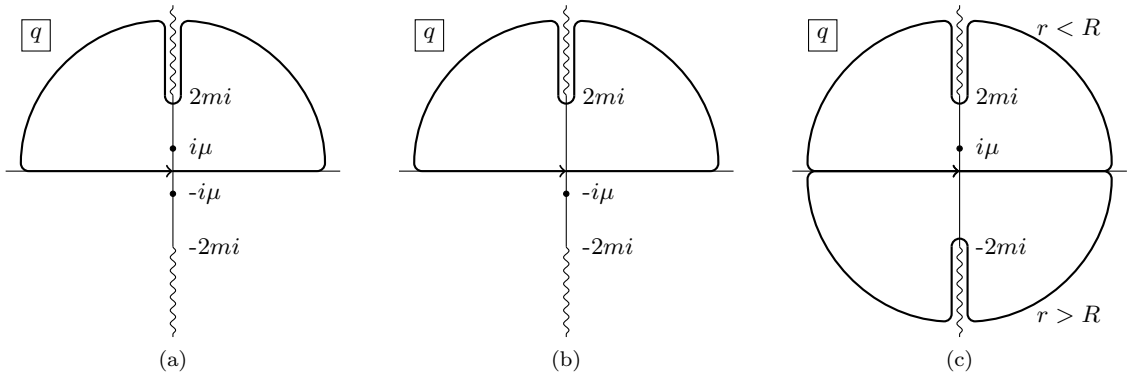

For a point-like source the Coulomb potential and its Fourier-transform $\tilde{\Phi}(q)$ are given by
\begin{equation}\label{eq:8}
\Phi(r) = -\frac{Ze}{4\pi r} \, e^{-\m r}\, , \quad \tilde{\Phi}(q)=-\frac{Ze}{q^2+\m^2},
\end{equation}
where the photon mass $\m$ is introduced for regularization of the integral in~\eqref{eq:6}. In this case for the function $c(r)$ one obtains the following expression
\begin{equation}\label{eq:9}
c(r)=\frac{1}{i} \int\limits_{-\infty}^\infty \! q dq \ e^{i q r} \(\frac{1}{q^2+\m^2}\) \frac{1}{\pi} \frac{F_2(-q^2)}{F_2(0)}\, .
\end{equation}
In~\eqref{eq:9} in the upper half-plane the integrand has a pole at the point $ i\m$, and besides, the electronic formfactor  $F_2(-q^2)$ has a cut on the imaginary axis, starting at $q=2mi$. The jump by transition through the cut amounts to $\Delta F_2(-q^2)=2 i \,\text{Im}\, F_2(-q^2)$. As a result, upon reducing the integral to the contour one (see Fig.~\ref{pic:1a}), the expression~\eqref{eq:9} in the limit $\m \to 0$ takes the form~\cite{Lautrup1976}
\begin{equation}\label{eq:10}
c(r)=1- \int\limits_{4m^2}^\infty \! \frac{dQ^2}{Q^2} \, e^{- Q r}\, \frac{1}{\pi}\, \frac{\text{Im}\, F_2(Q^2)}{F_2(0)} \, ,
\end{equation}
where $ \dfrac{1}{\pi}\, \text{Im}\, F_2(Q^2) = 2 F_2(0)\, \dfrac{m^2}{Q^2} \, \dfrac{1}{\sqrt{1-4m^2/Q^2}}$.

\subsection{The extended source\label{sec:2b}}

 For an extended nucleus the Coulomb potential, regularized by the same factor  $e^{-\m r}$ as in~\eqref{eq:8}, is taken in the standard form of a uniformly charged ball
 \begin{equation}\label{eq:11}
\Phi(r)=\left\lbrace\begin{aligned}
&-\dfrac{Z e}{4\pi r}\, e^{-\m r}\, , & r>R\\
&-\dfrac{Z e}{4\pi R}\,\dfrac{3R^2-r^2}{2R^2}\,e^{-\m r}\, , & r<R
\end{aligned}\right.\, ,
\end{equation}
where the nucleus radius $R$ is defined via  $Z$ by means of the  (simplified) expression $R(Z)\hm=1.228935 \  (2.5 Z)^\myfrac{1}{3} $~fm\footnote{The numerical coefficient in this expression is chosen in such a way, which provides the level $1s_{1/2}$ in the purely Coulomb problem with the potential $\F(r)$ and $Z\hm=170$ to lye very close to the threshold of negative continuum  with the bound energy $\simeq 1.99999\, m$. }. Then the Fourier-transform of the potential~\eqref{eq:11} can be represented as following
\begin{equation}\label{eq:12}
\tilde{\Phi}(q)=-\dfrac{Ze}{q}\(\tilde\Phi^{(+)}(q) e^{i q R} + \tilde{\Phi}^{(-)}(q) e^{-i q R} + \tilde{\Phi}^{(0)}(q) \),
\end{equation}
where
\begin{gather}\label{eq:13}
\tilde{\Phi}^{(+)}(q)=-3\,\frac{q R+i (1+R \mu )}{2 R^3 (q+i \mu )^4}\, e^{-R \mu }\, , \\
\tilde{\Phi}^{(-)}(q)=-3\,\frac{q R-i (1+R \mu )}{2 R^3 (q-i \mu )^4}\, e^{-R \mu }\, , \\
\tilde{\Phi}^{(0)}(q)=3 q \mu  \, \frac{q^4 R^2+\mu ^2 \left(-4+R^2 \mu ^2\right)+2 q^2 \left(2+R^2 \mu ^2\right)}{R^3 \left(q^2+\mu ^2\right)^4} \, .
\end{gather}
After substituting~\eqref{eq:12} into~\eqref{eq:6} one obtains
\begin{equation}
c(r)= J^{(+)}(r)+J^{(-)}(r)+J^{(0)}(r)
\end{equation}
where
\begin{subequations}\label{eq:14}
\begin{align}
J^{(\pm)}&=\frac{1}{i} \int\limits_{-\infty}^\infty \! dq \ e^{i q (r\pm R)} \, \tilde{\Phi}^{(\pm)}(q)\,  \frac{1}{\pi} \frac{F_2(-q^2)}{F_2(0)}\, , \\
J^{(0)}&=\frac{1}{i} \int\limits_{-\infty}^\infty \! dq \ e^{i q r}\, \tilde{\Phi}^{(0)}(q) \, \frac{1}{\pi} \frac{F_2(-q^2)}{F_2(0)}\, .
\end{align}
\end{subequations}

The integral $J^{(-)}$ is calculated differently in the cases $r<R$ and $r>R$, while the integrals  $J^{(+)}$, $J^{(0)}$ have the same form on the whole half-axis $r\in (0,\infty)$. In the integral $J^{(0)}$ the integration contour has the same form~\ref{pic:1a} as in the case of the point source, so the contribution comes from the pole $i\m$ only, since $\tilde{\Phi}^{(0)}\propto \m$, while the contribution from the cut vanishes upon the regularization removal $\m \rightarrow 0$. The Fourier-transform $\tilde{\Phi}^{(+)}$ has no poles in the upper half-plane, therefore the integral $J^{(+)}$ coincides with the integral along the cut of the function $F_2(-q^2)$ from the jump of the integrand by transition from the one side of the cut to another (see Fig.~\ref{pic:1b}). For $r<R$ in $J^{(-)}$ there remains the contribution from the cut $q\in(-2mi,-i\infty)$ only, since $\tilde{\Phi}^{(-)}$ doesn't possess any poles in the lower half-plane, while for $r>R$ the integral $J^{(-)}$  contains contributions from the pole $i\m$ and from the cut $q\in(2mi,i\infty)$ (see Fig.~\ref{pic:1c}). So the resulting expressions for the integrals~\eqref{eq:14} are
\begin{align}\label{eq:15}
J^{(0)}(r)&=\dfrac{1}{2 R^3}\left\lbrace r(3 R^2 - r^2) +3 i (r^2-R^2)\frac{ F_2'(0)}{F_2(0)}+3 r \frac{F_2''(0)}{F_2(0)}-i \frac{F_2'''(0)}{F_2(0)} \right\rbrace\, ,\\
J^{(+)}(r)&=-\int\limits_{4m^2}^\infty \! \frac{dQ^2}{Q^2}\, \frac{3(QR+1)}{R^3 Q^3}\, e^{- Q (r+R)}\, \frac{1}{\pi}\, \frac{\text{Im}\, F_2(Q^2)}{F_2(0)}\, , \\
J^{(-)}(r)&=\dfrac{1}{2 R^3}\left\lbrace (r-R)^2 (r+2 R) -3 i (r^2-R^2)\frac{ F_2'(0)}{F_2(0)}-3 r \frac{F_2''(0)}{F_2(0)}+i \frac{F_2'''(0)}{F_2(0)} \right\rbrace\ - {}\nonumber \\
{}&- \int\limits_{4m^2}^\infty \! \frac{dQ^2}{Q^2}\, \frac{3(QR-1)}{R^3 Q^3}\, e^{- Q (r-R)}\, \frac{1}{\pi}\, \frac{\text{Im}\, F_2(Q^2)}{F_2(0)}\, ,\qquad r>R\, , \\
J^{(-)}(r)&= \int\limits_{4m^2}^\infty \! \frac{dQ^2}{Q^2}\, \frac{3(QR+1)}{R^3 Q^3}\, e^{- Q (R-r)}\, \frac{1}{\pi}\, \frac{\text{Im}\, F_2(Q^2)}{F_2(0)}\, ,\qquad r<R \, ,
\end{align}
which by means of $F_2'(0)=0$, $F_2''(0)=- F_2(0)/3m^2$, $F_2'''(0)=0$ yield the final expression for the function $c(r)$ in the case of the uniformly charged extended nucleus~\eqref{eq:11}:
\begin{subequations}\label{eq:16}
\begin{align}
c(r)&=1- \int\limits_{4m^2}^\infty \! \frac{dQ^2}{Q^2}\, \frac{3QR\ch QR - 3\sh QR}{R^3 Q^3} \, e^{- Q r}\, \frac{1}{\pi}\, \frac{\text{Im}\, F_2(Q^2)}{F_2(0)}, \quad r>R\, ,\\
c(r)&=\frac{(3 R^2 - r^2)}{2 R^3}\,r - \frac{r}{2 m^2 R^3} +{} \nonumber\\
{}&+ \int\limits_{4m^2}^\infty \! \frac{dQ^2}{Q^2}\, \frac{3(QR+1)}{R^3 Q^3}\,\sinh Qr \, e^{- Q R}\, \frac{1}{\pi}\, \frac{\text{Im}\, F_2(Q^2)}{F_2(0)} , \qquad \qquad \qquad r<R\, .
\end{align}
\end{subequations}

And although the function  $c(r)$ by itself doesn't possess any transparent physical sense, the expression $\D g_{free}\, c(r)$ could be interpreted as the dependence of the electronic AMM on the distance from the nucleus center. The inspection of expressions~(\ref{eq:10},\ref{eq:16}) shows, that $c(r) \to  1$ in the region $r\gtrsim 1/m$, while for $r \to 0$ it monotonically tends to zero. Thus, regardless the magnitude of the charge of the Coulomb source the behavior of the effective potential~\eqref{eq:7} turns out to be substantially different from~\eqref{eq:1} first of all at small distances $\ll 1/m$, whereas in the low-energy limit, i.e. at the distances, exceeding $1/m$, both potentials are quite close.

\section{The Dirac equation with $\Delta U_{AMM}$\label{sec:3}}

The general form of DE for an electron in the Coulomb field of the nucleus with account for  the additional  effective interaction due to AMM~\eqref{eq:7} takes the form ($\h\hm=c\hm=m\hm=1$)
\beq \label{eq:18}
\(\vec{\a}\vec{p}+\b  +U(r)  + \D U_{AMM} \)\p=E\psi \ ,
\eeq
where $U(r) = e \Phi (r)$ is the Coulomb interaction, in which the quark structure doesn't play any special role, and so  $\Phi(r)$ could be taken in the form~\eqref{eq:11}. In dependence on the source under consideration for $\Delta U_{AMM}$ (the extended nuclei with the charge $Z$ as a whole or separate quarks inside the nucleus with the charge $Z_q$) the function  $c(r)$, defined in~\eqref{eq:16} (denoted as $c_N(r)$) or in~\eqref{eq:10} (denoted as $c_q(r)$) will be used.
Then for the upper $i \vf $ and the lower $\c$ components of the Dirac bispinor there follows from the eq.~\eqref{eq:18}
\beqar \label{eq:19}
i \( \v {\s}\v p + \l \[\v {\s}\v p \ , V (\v {r} \,)\]\) \vf & = &\(\e+1-U(r)\)\c \ ,  \nonumber \\
i \(\v {\s}\v p - \l \[\v {\s}\v p \ , V (\v {r} \,) \]\) \c & = & -\(\e-1-U(r)\) \vf \ ,
\eeqar
  wherein $V(\vec{r}\,)=Z c_N(r)/r$ for the nucleus, considered as a separate uniformly charged Coulomb source for $\Delta U_{AMM}$, while with account for the quark structure $V(\vec{r}\,)$ should be compiled via  $c_q(\v r)$ found from  the concrete quark configuration.

As a first step we  consider the central problem. The main motivation to such approximation is that the nucleus size is much less compared to the mean radius of the electronic WF, and so to the leading order the displacement of quarks from the nucleus center could be neglected. In this case for a separate quark $q$ in the nucleus center $V(\vec{r}\,)=Z_q c_q(r)/r$.

Both for central quarks and for the whole nucleus in presence of  $\D U_{AMM}$ the total moment of the electron $\vec{j}$ and the operator $k=\b( \vec \s \ \vec l +1) \ $ are still conserved, hence, in the standard representation for the Dirac matrices the upper and lower components of the electronic WF will contain the spherical spinors $\W_{jlm_j}$ and $ \W_{jl'm_j}$   of different parity, $l+l'=2j$, and the real radial functions $f_j(r)$  and $g_j(r)$ correspondingly
\beq \label{eq:20} \psi_{jm_j}=\frac{1}{ r} \left(
\begin{array}{c}
 i f_j(r)\, \W_{jlm_j} \\
 g_j(r)\, \W_{jl'm_j}
 \end{array}
\right) \ , \eeq
The definition of spherical harmonics and spinors follows~\cite{Bateman1953}, whence $\W_{jl'm_j}\hm=(\vec{\s} \vec{n})\, \W_{jlm_j}$. The states with $j=l\hm+\myfrac{1}{2}$ and different parity are distinguished  via different values of $\k=\pm (j+\myfrac{1}{2})$ and are subject of equations
\beqar \label{eq:21}
   \( \pd_{r}- Z\l\nu(r)/r^2 + \k/r \)f_j &=& (\e+1-U(r))g_j   \ , \nonumber \\
   \(\pd_{r} + Z\l\nu(r)/r^2 - \k/r \)g_j &=& -(\e-1-U(r))f_j \ , \quad
\eeqar
where $\nu(r)=c(r)-r c'(r)$.

\begin{figure}
\subfigure[]{
\includegraphics[width=.48\textwidth]{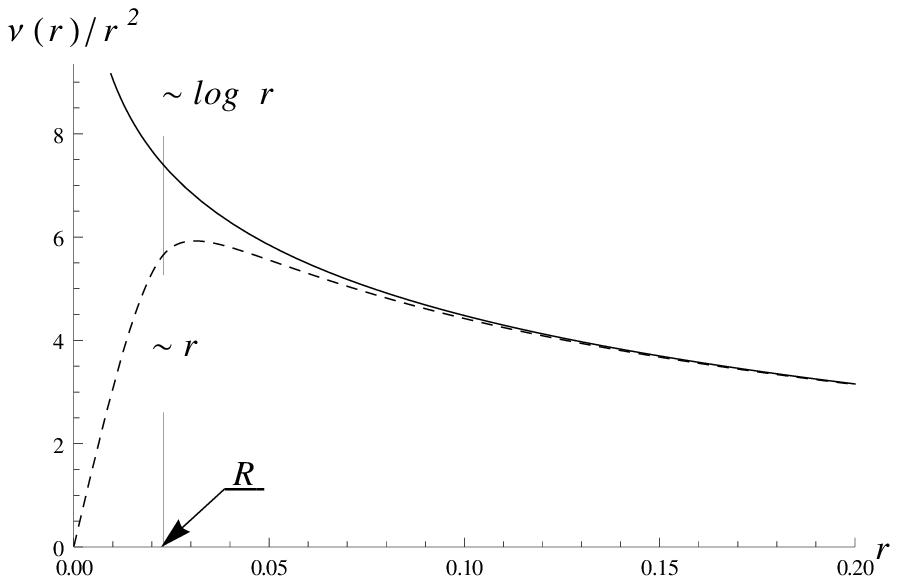} \label{pic:2a}
}
\hfill
\subfigure[]{
\includegraphics[width=.48\textwidth]{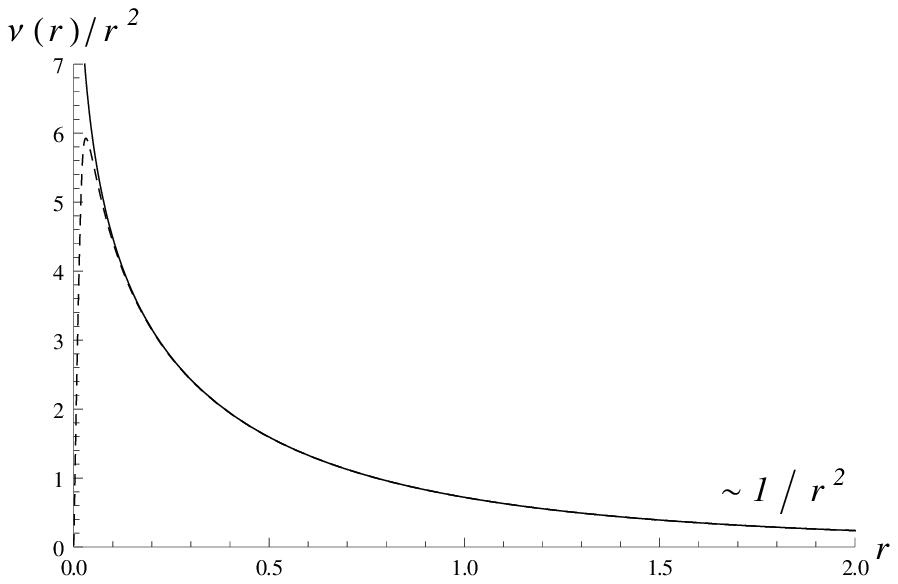} \label{pic:2b}
}
\caption{The behavior of the term $\nu(r)/r^2$ in eqs.~\eqref{eq:21} at small~\subref{pic:2a} and large~\subref{pic:2b} distances for a point-like quark in the nucleus center (solid line) and for an extended nucleus with radius $R$ (dotted line).}\label{pic:2}
\end{figure}

Here it should be noted, that in the case of central quarks the term $\nu_q(r)/r^2$ behaves for $r \ll 1 $ as $\log r$ (corresponding to $c(r) \to 0$), while for $r \to \inf$ it behaves as $1/r^2$ (what corresponds to $c(r)\to~1$)~\cite{Lautrup1976}. For an extended nucleus the difference in behavior of $\nu_N(r)/r^2$ compared to the point-like source arises at the scales of nucleus size, moreover,  for $r\ll R$ the term $\nu_N(r)/r^2$ behaves almost linearly and vanishes in the nucleus center (see Fig.~\ref{pic:2}). The latter shows once more that the main role in the dynamical screening of AMM is played by small distances, rather than by the external field magnitude, which also vanishes in the nucleus center.

The numerical solution of the system~\eqref{eq:21} for the central quarks with $c(r)\hm=c_q(r)$ shows, that for the  nucleus with the critical charge $Z=170$  the level $1s_{1/2}$, corresponding to $\k=-1$,  is shifted relative to the purely Coulomb case with the potential~\eqref{eq:11} in the next way
\begin{subequations}\label{eq:23}
\beq \label{eq:23a}
\D \e_{u}(1s_{1/2})=4.58 \text{\, eV} \, ,  \quad  \D \e_{d}(1s_{1/2})=-2.29 \text{\, eV} \, ,
\eeq
while for the nucleus with $Z=183$ and the level $2p_{1/2}$ ($\k=1$) the similar calculation gives
\beq
\D \e_{u}(2p_{1/2})=-4.29 \text{\, eV} \, ,  \quad \D \e_{d}(2p_{1/2})=2.15 \text{\, eV}\, .\label{eq:23b}
\eeq
\end{subequations}
It should be specially noted here, that the possibility of calculation of downward energy shifts  relative to the purely Coulomb case is provided by such a choice of the dependence $R(Z)$ in~\eqref{eq:11}, when the lowest Coulomb levels $1s_{1/2}$ for $Z=170$ and $2p_{1/2}$ for $Z=183$ lye a little higher than the threshold of the lower continuum.

These results for the contribution from $\D U_{AMM}$ to the levels shift are completely non-perturbative in $Z\a$  and  (partially) in $\a/\pi$. The reason for the latter circumstance is that the actual coupling constant in  $\D U_{AMM}$ is  $Z_q\l$, in which $\a/\pi$ enters as a cofactor. It should be stressed, however, that this dependence has nothing to do with the  summation of the loop expansion for AMM, since the initial expression for the operator~\eqref{eq:6} is based on the one-loop approximation for the vertex. At the same time, since within the one-particle DE~\eqref{eq:18} there exists a possibility   of non-perturbative evaluation of the  contribution from $\D U_{AMM}$, this should be used in order to compare the results of perturbative and non-perturbative approaches to  $\D U_{AMM}$ at large $Z$, in view of recent essentially non-perturbative calculations of the vacuum energy for $Z > Z_{cr,1}$~\cite{Davydov2017, Voronina2017}, which show an explicitly nonlinear nature of this effect outside PT.

At first, the results~\eqref{eq:23} should be compared with the estimate for the shift via PT, when  $\D U_{AMM}$ is considered as a perturbation of the Coulomb potential~\eqref{eq:11}. In this case the shift of Coulomb levels with quantum numbers $nj$, caused by central quarks, is found by means of the expression
\begin{equation}\label{eq:24}
\Delta \e_q(nj)^{\text{PT}} = \langle \psi^{(0)}_{nj} | \Delta U_{AMM} | \psi^{(0)}_{nj} \rangle = - 2Z_q\l\int\limits_{0}^{\infty} \! dr f^{(0)}_{nj}(r)\, g^{(0)}_{nj}(r)\, \nu_q(r)\, ,
\end{equation}
with $f_{nj}^{(0)}(r)\, , \ g_{nj}^{(0)}(r)$ being the radial components of the unperturbed   Coulomb level with definite parity $\p_{nj}^{(0)}$.  Calculated in accordance with~\eqref{eq:24} the values $\D \e_q^{\text{PT}}$ coincide with results~\eqref{eq:23} with precision not less than $0.1\%$.
Taking account of that in the case under consideration the mean radius of the electronic WF should be  $O(1) \gg R(Z)$,  the probability for an electron staying inside the nucleus  $\simeq 0.02$, both with quarks in the nucleus center and without them,  than at first glance the contribution of quarks shifted from the nucleus center may differ from~\eqref{eq:23} only in the amendment. So to the leading order the resulting shift of the electronic level could be estimated via direct sum
\beq \label{eq:25}
\D \e_{AMM}=(2Z+N) \D \e_{u} + (Z+2N) \D \e_{d}\ ,
\eeq
with  $Z$ being the number of protons, $N \simeq 1.5 Z$ --- the number of neutrons (for large  $Z$). As a result, for the shift caused by  $\D U_{AMM}$ of the levels $1s_{1/2}$  for the nucleus with  $Z=170$  and  $2p_{1/2}$ for $Z=183$ one finds
\beq \label{eq:26}
\D \e_{AMM}\,(1s_{1/2}, Z=170)=1.17 \text{\, KeV} \, , \quad \D \e_{AMM}\,(2p_{1/2}, Z=183)=-1.18 \text{\, KeV} \, .
\eeq

At the same time, when the quark structure of the nucleus is ignored and so $c_N(r)$ defined in~\eqref{eq:16} is used, then for the total shift of levels one obtains by means of numerical solution of the system~\eqref{eq:21} for the whole nucleus as the source for $\D U_{AMM}$
\beq \label{eq:28}
\mathcal{D} \e_{AMM}\,(1s_{1/2}, Z=170)=1.12 \text{\, KeV} \, , \quad \mathcal{D} \e_{AMM}\,(2p_{1/2}, Z=183)=-1.09 \text{\, KeV} \, .
\eeq
The estimate via PT in this case could be found via Coulomb WF $\p_{nj}^{(0)}$ by means of expression
\begin{equation}\label{eq:29}
\mathcal{D} \e\,(nj)^{\text{PT}} = \langle \psi^{(0)}_{nj} | \Delta U_{AMM} | \psi^{(0)}_{nj} \rangle =- 2Z\l\int\limits_{0}^{\infty} \! dr f^{(0)}_{nj}(r) g^{(0)}_{nj}(r)\nu_N(r)\, ,
\end{equation}
and gives the results, coinciding with the  non-perturbative ones~\eqref{eq:28} with precision not less than $0.1\%$, as well as the perturbative and non-perturbative results for the shifts from separate central quarks~(\ref{eq:23},\ref{eq:24}). However, in the central quarks approximation the results for the level shifts, calculated via quark structure~\eqref{eq:26} and the extended uniformly charged nucleus~\eqref{eq:28}, turn out to be noticeably  different, and in what follows we'll show, how by taking account of quarks residing in the nucleus periphery this correspondence could be significantly improved.

\section{Contribution from the periphery as a perturbation\label{sec:4}}

The results~\eqref{eq:26} for the shifts of the levels $1s_{1/2}$ and $2p_{1/2}$ are obtained  under  assumption, that all the quarks in the nucleus can be considered as the central ones. However, the remarkable difference between these results with~\eqref{eq:28} points out, that the contribution from the quarks residing in the nucleus periphery should be considered separately.

At first, let us consider the displacement of quarks from the nucleus center as a perturbation.
For these purposes within PT one should firstly find the correction, caused by the displacement   $\vec{a}$ of the central quark, and thereafter take the average of this correction over all the vectors subject to condition  $|\vec{a}|\leq R$.

Let us start with the case, when the initial approximation  corresponds to the quarks in the center, hence, the unperturbed WF should contain via Coulomb term in~\eqref{eq:18} the complete dependence on $Z\a$ and simultaneously  depend nonlinearly on $\a/\pi$ via factor $Z_q \l$ in $\D U_{AMM}$. Then the total shift of the level with quantum numbers $nj$ with account for averaging the quark position over the nucleus volume amounts to $\Delta \e_{q} + \d\e_q$, where
\begin{equation}\label{eq:31}
\d \e_q (nj) =\frac{3}{4\pi R^3}\int\!d\vec{a}\ \< \p^{(q)}_{nj}|\d U_{AMM}(\vec{a})|\p_{nj}^{(q)}\> \, ,
\end{equation}
with   $\p_{nj}^{(q)}, \ f_{nj}^{(q)}, \ g_{nj}^{(q)}$ being the WF of the problem~\eqref{eq:21} with the quark $q$ in the nucleus center, while
\begin{equation}\label{eq:32}
\d U_{AMM}(\vec{a})=-i\,Z_q\lambda\, \( \nu_q(|\vec{r}-\vec{a}|)\frac{\vec{\gamma}\cdot(\vec{r}-\vec{a})}{|\vec{r}-\vec{a}|^3} - \nu_q(r)\frac{\vec{\gamma}\cdot\vec{r}}{r^3} \)
\, .
\end{equation}
After some algebra the expression~\eqref{eq:31} takes the following form  (for the levels $1s_{1/2}$ and $2p_{1/2}$)
\begin{multline}\label{eq:33}
\d \e_q(nj) = 2Z_q\l\int\limits_{0}^{\infty} \! dr f^{(q)}_{nj}(r)\, g^{(q)}_{nj}(r)\, \nu_q(r) - {}\\
{}-Z_q \l \frac{3}{R^3} \! \int\limits_0^R \!a^2da \! \int\limits_0^\infty \! r^2dr \! \int\limits_{-1}^1\! dx \ f^{(q)}_{nj}(r) g^{(q)}_{nj}(r) \, \nu_q\big([r^2+a^2-2rax]^\myfrac{1}{2}\big)\,\frac{r-ax}{[r^2+a^2-2rax]^\myfrac{3}{2}}\, .
\end{multline}
The total shift of levels due to $\D U_{AMM}$  with account for averaging over the nucleus volume is determined quite similar to \eqref{eq:25}
\beq \label{eq:55}
D \e_{AMM}=(2Z+N) (\D \e_{u} + \d\e_u) + (Z+2N)(\D \e_{d} + \d\e_d) \, .
\eeq

\begin{figure}
\center
\includegraphics[width=.49\textwidth]{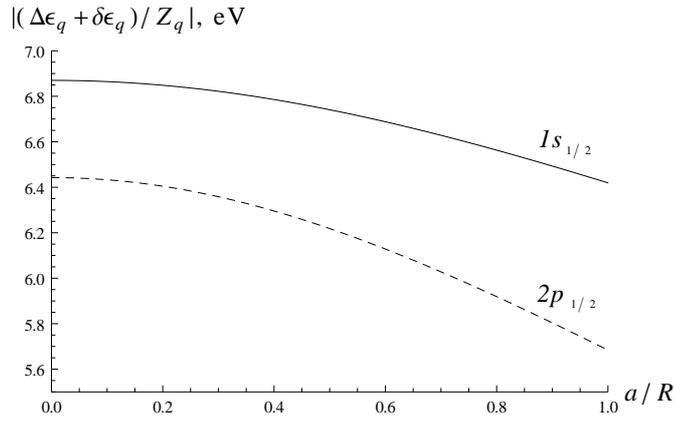}
\caption{The dependence $|(\D \e_q +\d \e_{q}(a))/Z_q|$ is given in eV. The level $1s_{1/2}$ for $Z=170$ is shown by the solid line, while the level $2p_{1/2}$ for $Z=183$  --- by the dotted one. For the positive charge of the source the shift of the level $1s_{1/2}$ is positive, while the shift of the level $2p_{1/2}$ is negative.}\label{pic:5}
\end{figure}

Before presenting the results of calculations, it should be noted, that in~\eqref{eq:31} it is possible to use as the initial approximation the purely Coulomb functions, quite similar to the case of estimates~\eqref{eq:24},~\eqref{eq:29}. Then, as it was already mentioned in the Section~\ref{sec:3}, the obtained results should contain the complete dependence on $Z\a$, but at the same time represent an effect of the first order in $\a/\pi$. In this case for the shift of levels arising from the interaction  via AMM~\eqref{eq:7} of the electron with the quark shifted by distance $a$ from the nucleus center, one gets (for the levels $1s_{1/2}$ and $2p_{1/2}$):
\begin{multline}\label{eq:56}
(\D \e_q +\d \e_{q})^{\text{PT}}(a,nj) =\langle \psi^{(0)}_{nj} | -i\,Z_q\lambda\,\(\vec{\gamma}\cdot(\vec{r}-\vec{a})\)\frac{\nu_q(|\vec{r}-\vec{a}|)}{|\vec{r}-\vec{a}|^3} | \psi^{(0)}_{nj} \rangle = \\
=-Z_q \l \! \int\limits_0^\infty \! r^2dr \! \int\limits_{-1}^1\! dx \ f^{(0)}_{nj}(r) g^{(0)}_{nj}(r) \, \nu_q\big([r^2+a^2-2rax]^\myfrac{1}{2}\big)\,\frac{r-ax}{[r^2+a^2-2rax]^\myfrac{3}{2}}\, .
\end{multline}
The dependence of $|(\D \e_q +\d \e_{q}(a))^{\text{PT}}/Z_q|$ on the displacement $a$ is shown in the Fig.~\ref{pic:5}  for the levels $1s_{1/2}$ and $2p_{1/2}$ with $Z\sim Z_{cr}$.

Proceeding further, from~\eqref{eq:56}  one finds the shift, averaged over the position inside the nucleus, which is caused by one quark of the given type $q$
\beq\label{eq:57}
(\D \e_q +\d \e_{q})^{\text{PT}}(nj) =\frac{3}{4\pi R^3}\int\!d\vec{a}\ (\D \e_q +\d \e_{q})^{\text{PT}}(a,nj)\, ,
\eeq
whereupon similarly to~\eqref{eq:25} one finds the total shift $D\e^{\text{PT}}(nj)$.

In Tabs.~\ref{tab:1b},~\ref{tab:2b} there are presented  the shifts of levels $1s_{\myfrac{1}{2}}$ and $2p_{\myfrac{1}{2}}$ caused by AMM  for separate quarks, which are found via different methods, including non-perturbative solution of the system~\eqref{eq:21} with central quarks, and also using PT (\ref{eq:24}), as well as with account for the spatial displacement (\ref{eq:31},~\ref{eq:57}). Quite similar to the preceding Section, the shifts, found within PT with unperturbed Coulomb WF, are marked by the label ,,${\text{PT}}$''. It should be noted, that with account for displacement the shifts $\D \e_q +\d \e_{q}$ and $(\D \e_q +\d \e_{q})^{\text{PT}}$, calculated with different degree of accuracy in $\a/\pi$,  differ from each other by no more than $~0.1\%$ for both types of quarks, and so the corresponding curves for $|(\D \e_q +\d \e_{q}(a))/Z_q|$, shown in Fig.~\ref{pic:5}, are actually indistiguishable.

\begin{table}
\caption{The shifts for separate quarks (in eV) for  $1s_{1/2}$. \label{tab:1b}}
\begin{ruledtabular}
\begin{tabular*}{1.0\textwidth}{@{\extracolsep{\fill} } b{3.0} *{2}{b{2.4}} b{2.6} *{2}{b{2.4}} }
\mc{ $Z$ } & \mc { $\Delta \epsilon _u$ } & \mc { $\Delta \epsilon _u^{\text{PT}} $} & \mc { $\d \e_u $} & \mc { $\Delta \e_u + \d \e_u $} & \mc { $(\Delta \e_u + \d \e_u)^{\text{PT}} $ }\\
\colrule\noalign{\smallskip}
 80 & 0.3637 & 0.3644 & -0.000228 & 0.3635 & 0.3640  \\
 90 & 0.5157 & 0.5168 & -0.000490 & 0.5152 & 0.5160 \\
 100 & 0.7140 & 0.7152 & -0.00104 & 0.7130 & 0.7135 \\
 110 & 0.9741 & 0.9753 & -0.00221 & 0.9719 & 0.9726 \\
 120 & 1.318 & 1.320 & -0.00480 & 1.314 & 1.316 \\
 130 & 1.777 & 1.779 & -0.0106 & 1.767 & 1.770  \\
 140 & 2.380 & 2.382 & -0.0238 & 2.356 & 2.360\\
 150 & 3.120 & 3.125 & -0.0522 & 3.068 & 3.072 \\
 160 & 3.905 & 3.910 & -0.106 & 3.799 & 3.804  \\
 170 & 4.575 & 4.580 & -0.190 & 4.385 & 4.390  \\
\colrule\noalign{\smallskip}
\mc{ $ Z$ } & \mc {  $\Delta \epsilon _d$ } & \mc { $\Delta \epsilon _d^{\text{PT}} $} & \mc { $\d \e_d $} & \mc { $\Delta \e_d + \d \e_d $} & \mc { $(\Delta \e_d + \d \e_d)^{\text{PT}} $}\\
\colrule\noalign{\smallskip}
 80 &  -0.1819 & -0.1822 & 0.000114 & -0.1817 & -0.1820 \\
 90 &  -0.2579 & -0.2584 & 0.000245 & -0.2576 & -0.2580 \\
 100 &  -0.3570 & -0.3576 & 0.000520 & -0.3565 & -0.3568 \\
 110 &  -0.4871 & -0.4876 & 0.00111 & -0.4860 & -0.4863 \\
 120 &  -0.6592 & -0.6601 & 0.00240 & -0.6568 & -0.6578 \\
 130 &  -0.8887 & -0.8897 & 0.00532 & -0.8834 & -0.8849 \\
 140 &  -1.190 & -1.191 & 0.0119 & -1.178 & -1.180 \\
 150 &  -1.560 & -1.563 & 0.0261 & -1.534 & -1.536 \\
 160 &  -1.953 & -1.955 & 0.0529 & -1.900 & -1.902 \\
 170 &  -2.287 & -2.290 & 0.0950 & -2.192 & -2.195 \\
\end{tabular*}
\end{ruledtabular}
\end{table}

\begin{table}
\caption{The shifts for separate quarks (in eV) for $2p_{1/2}$.\label{tab:2b}}
\begin{ruledtabular}
\begin{tabular*}{1.\textwidth}{@{\extracolsep{\fill} } c *{2}{b{2.5}} b{2.7} *{2}{b{2.5}} }
\mc{ $ Z $} & \mc {  $\Delta \epsilon _u $} & \mc { $\Delta \epsilon _u^{\text{PT}} $} & \mc { $\d \e_u $} & \mc { $\Delta \e_u + \d \e_u $} & \mc { $(\Delta \e_u + \d \e_u)^{\text{PT}} $ }\\
\colrule\noalign{\smallskip}
 90 & -0.06067 & -0.06066 & 0.0000154 & -0.06069 & -0.06067 \\
 100 & -0.09808 & -0.09807 & 0.0000450 & -0.09813 & -0.09813 \\
 110 & -0.1599 & -0.1599 & 0.000138 & -0.1601 & -0.1601 \\
 120 & -0.2680 & -0.2680 & 0.000453 & -0.2688 & -0.2689 \\
 130 & -0.4717 & -0.4718 & 0.00166 & -0.4750 & -0.4751 \\
 140 & -0.8793 & -0.8794 & 0.00665 & -0.8901 & -0.8903 \\
 150 & -1.636 & -1.637 & 0.0260 & -1.654 & -1.655 \\
 160 & -2.652 & -2.654 & 0.0782 & -2.633 & -2.637 \\
 170 & -3.558 & -3.560 & 0.169 & -3.431 & -3.433 \\
 180 & -4.171 & -4.174 & 0.285 & -3.896 & -3.897 \\
 183 & -4.293 & -4.295 & 0.322 & -3.970 & -3.972 \\
\colrule\noalign{\smallskip}
\mc{ $ Z $} & \mc {  $\Delta \epsilon _d $} & \mc { $\Delta \epsilon _d^{\text{PT}} $} & \mc { $\d \e_d $} & \mc { $\Delta \e_d + \d \e_d $} & \mc { $(\Delta \e_d + \d \e_d)^{\text{PT}} $ }\\
\colrule\noalign{\smallskip}
 90 &   0.03035 & 0.03034 & -0.00000768 & 0.03034 & 0.03034 \\
 100 &  0.04909 & 0.04909 & -0.0000225 & 0.04907 & 0.04906 \\
 110 &  0.08013 & 0.08014 & -0.0000688 & 0.08006 & 0.08007 \\
 120 &  0.1346 & 0.1347 & -0.000227 & 0.1344 & 0.1345 \\
 130 &  0.2383 & 0.2384 & -0.000828 & 0.2375 & 0.2375 \\
 140 &  0.4484 & 0.4485 & -0.00333 & 0.4451 & 0.4452 \\
 150 &  0.8402 & 0.8406 & -0.0130 & 0.8272 & 0.8276 \\
 160 &  1.355 & 1.358 & -0.0391 & 1.316 & 1.318 \\
 170 &  1.800 & 1.801 & -0.0844 & 1.715 & 1.716 \\
 180 &  2.090 & 2.092 & -0.142 & 1.948 & 1.949 \\
 183 &  2.146 & 2.147 & -0.161 & 1.985 & 1.986 \\
\end{tabular*}
\end{ruledtabular}
\end{table}

\begin{table}
\caption{The total shifts for $1s_{1/2}$  (in KeV).\label{tab:1}}
\begin{ruledtabular}
\begin{tabular*}{1.0\textwidth}{@{\extracolsep{\fill} }b{3.0} *{6}{b{2.6}}}
\mc{$Z$ } & \mc { $\Delta \epsilon $} & \mc { $\Delta \epsilon^{\text{PT}} $} & \mc { $\mathcal{D} \epsilon $} & \mc { $D\epsilon $} & \mc { $\mathcal{D}\epsilon^{\text{PT}} $} & \mc { $D \epsilon^{\text{PT}} $}\\
\colrule\noalign{\smallskip}
 80 & 0.043646 & 0.043724 & 0.043597 & 0.043618 & 0.043697 & 0.043676 \\
 90 & 0.069619 & 0.069762 & 0.069515 & 0.069552 & 0.069696 & 0.069658 \\
 100 & 0.107106 & 0.107284 & 0.106887 & 0.106950 & 0.107128 & 0.107030 \\
 110 & 0.160729 & 0.160921 & 0.160265 & 0.160364 & 0.160556 & 0.160482 \\
 120 & 0.237322 & 0.237618 & 0.236310 & 0.236458 & 0.236753 & 0.236791 \\
 130 & 0.346579 & 0.346967 & 0.344301 & 0.344506 & 0.344890 & 0.345100 \\
 140 & 0.499771 & 0.500275 & 0.494512 & 0.494765 & 0.495260 & 0.495551 \\
 150 & 0.701953 & 0.703130 & 0.689946 & 0.690200 & 0.691335 & 0.691157 \\
 160 & 0.937248 & 0.938431 & 0.911676 & 0.911853 & 0.912975 & 0.912911 \\
 170 & 1.166570 & 1.167942 & 1.118285 & 1.118136 & 1.119560 & 1.119541 \\
\end{tabular*}
\end{ruledtabular}
\end{table}

\begin{table}
\caption{The total shifts for $2p_{1/2}$ (in KeV).\label{tab:2}}
\begin{ruledtabular}
\begin{tabular*}{1.0\textwidth}{@{\extracolsep{\fill} }b{3.0} *{6}{b{2.6}}}
\mc{$Z$ } & \mc { $\Delta \epsilon $} & \mc { $\Delta \epsilon^{\text{PT}} $} & \mc { $\mathcal{D} \epsilon $} & \mc { $D\epsilon $} & \mc { $\mathcal{D}\epsilon^{\text{PT}} $} & \mc { $D \epsilon^{\text{PT}} $}\\
\colrule\noalign{\smallskip}
 90 & -0.008186 & -0.008184 & -0.008197 & -0.008193 & -0.008191 & -0.008191 \\
 100 & -0.014691 & -0.014690 & -0.014730 & -0.014720 & -0.014719 & -0.014719 \\
 110 & -0.026302 & -0.026306 & -0.026443 & -0.026421 & -0.026424 & -0.026424 \\
 120 & -0.047912 & -0.047925 & -0.048442 & -0.048393 & -0.048409 & -0.048408 \\
 130 & -0.090675 & -0.090729 & -0.092737 & -0.092627 & -0.092640 & -0.092639 \\
 140 & -0.179753 & -0.179775 & -0.187155 & -0.186931 & -0.186966 & -0.186965 \\
 150 & -0.354971 & -0.355106 & -0.372568 & -0.372250 & -0.372442 & -0.372440 \\
 160 & -0.617588 & -0.617420 & -0.632095 & -0.631870 & -0.632793 & -0.632790 \\
 170 & -0.893154 & -0.893796 & -0.874857 & -0.874811 & -0.875325 & -0.875320 \\
 180 & -1.122856 & -1.123497 & -1.051722 & -1.051839 & -1.052329 & -1.052321 \\
 183 & -1.178294 & -1.178957 & -1.089671 & -1.089837 & -1.090388 & -1.090379 \\
\end{tabular*}
\end{ruledtabular}
\end{table}

In Tabs.~\ref{tab:1},~\ref{tab:2} the total shifts of the levels $1s_{1/2}$ and $2p_{1/2}$ due to AMM, calculated both non-perturbatively on the basis of solution of the system~\eqref{eq:21} for the whole nucleus~\eqref{eq:28} and via PT~\eqref{eq:29}, as well as based on results from Tabs.~\ref{tab:1b},~\ref{tab:2b}, are shown. As it follows from these results, accounting for the contribution of central quarks only (two first columns in Tab.~\ref{tab:1},~\ref{tab:2}) with increasing $Z$ begins to diverge quite seriously (already in the third digit) with more precise calculations, including the contribution from the nucleus periphery. In turn, the contribution from the nucleus periphery adjust the results, found for the whole nucleus as the Coulomb source for $\D U_{AMM}$~\eqref{eq:28},~\eqref{eq:29},  with those, obtained via summation of separate contributions from all the quarks in the nucleus~\eqref{eq:31},~\eqref{eq:57}, although the account for the periphery according to~\eqref{eq:28},~\eqref{eq:29} and to~\eqref{eq:31},~\eqref{eq:57} is substantially different --- in the first case the total charge is defined as a uniform density distribution over the nucleus volume, whereas in the second one the point-like quarks are shifted. The comparison of results found by means of different methods shows that the best coincidence is achieved for the estimates~\eqref{eq:28} and~\eqref{eq:55} (3,4 columns), obtained with non-perturbative account for both  $Z\a$ and  (partially)  $\a/\pi$, as well as~\eqref{eq:29} and the corresponding sum of separate shifts~\eqref{eq:57}, which are found by means  of PT using the non-perturbed Coulomb WF (5,6 columns). With growing of  $Z$ there appears also a difference between perturbative and non-perturbative results for the total shifts of $1s_{1/2}$ and $2p_{1/2}$, but only in the following (forth) digit.

\section{Non-perturbative calculation  of $\Delta U_{AMM}$ for several shifted quarks\label{sec:Bb}}

Up to now the contribution from separate quarks to the shift of electronic levels has been considered by taking into account their displacement from the nucleus center as a perturbation, ignoring their  mutual correlations. The non-perturbative evaluation of the interaction between the electron and several shifted from the nucleus center point-like sources (quarks) due to $\Delta U_{AMM}$ could be implemented in the next way. Let us consider firstly two quarks with the charge $Z_q$, placed on the $z$-axis at the points $z=\pm a$. Then
\begin{equation}
V(\vec{r}\,)=Z_q\(\frac{ c_q(|\vec{r}-\vec{a}|)}{|\vec{r}-\vec{a}|}+\frac{c_q(|\vec{r}+\vec{a}|)}{|\vec{r}+\vec{a}|}\)\, , \quad \vec{a}= a \vec{e}_z \ .
\end{equation}


The spinors $\varphi, \chi$, corresponding to the solution of the system~\eqref{eq:19} with definite parity and $m_j$, are seed now as expansions in spherical spinors, which for an even level take the form
\begin{subequations}\label{eq:62}
\begin{align}\label{eq:62ev}
\varphi &= \sum_{k=0} \left( u_k \, \Omega^{(+)}_{2k,m_j} + v_k \, \Omega^{(-)}_{2k+2,m_j} \right) ,  &
\chi &=\sum_{k=0} \left( p_k \, \Omega^{(+)}_{2k+1,m_j} + q_k \, \Omega^{(-)}_{2k+1,m_j} \right),
\end{align}
\text{while for an odd one}
\begin{align} \label{eq:62od}
\varphi &=  \sum_{k=0} \left(u_k \, \Omega^{(+)}_{2k+1,m_j} + v_k \, \Omega^{(-)}_{2k+1,m_j} \right) , &\chi &=\sum_{k=0}\left( p_k \, \Omega^{(+)}_{2k,m_j} + q_k \, \Omega^{(-)}_{2k+2,m_j} \right) ,
\end{align}
\end{subequations}
where the definitions $\Omega^{(+)}_{l,m_j}\equiv\Omega_{jlm_j}$ and $\Omega^{(-)}_{l+1,m_j}\hm\equiv(\vec{\sigma} \vec{n} )\, \Omega_{jlm_j}$, which are quite convenient in what follows, are introduced, while all the radial functions $u_k, v_k, p_k, q_k$ could be taken real.

The substitution of expansions (\ref{eq:62ev},\ref{eq:62od}) in eq.~\eqref{eq:19} leads to a system of equations for the radial functions $u_k, v_k, p_k, q_k$, its own for each parity. For an even level with fixed $m_j$ the corresponding system takes the form
\begin{subequations}\label{eq:63}
\begin{equation}\label{eq:63ev}
\begin{split}
	\partial_r u_{k} &- \frac{2k}{r}u_{k}  + \lambda \sum_{s} \Big( \mathcal{A}_{2k;2s}(r)u_{s} +  \mathcal{B}_{2k;2s+2}(r)v_{s}\Big) = (\epsilon +1 -U(r))q_{k} \\
	\partial_r v_{k} &+ \frac{2k+3}{r}v_{k} + \lambda \sum_{s} \Big( \mathcal{C}_{2k+2;2s}(r)u_{s} + \mathcal{D}_{2k+2;2s+2}(r)v_{s}\Big) = (\epsilon +1 -U(r))p_{k} \\
	\partial_r p_{k} &- \frac{2k+1}{r}p_{k} - \lambda \sum_{s} \Big( \mathcal{A}_{2k+1;2s+1}(r)p_{s} + \mathcal{B}_{2k+1;2s+1}(r)q_{s}\Big) = -(\epsilon -1 -U(r))v_{k} \\
	\partial_r q_{k} &+ \frac{2k+2}{r}q_{k} - \lambda \sum_{s} \Big( \mathcal{C}_{2k+1;2s+1}(r)p_{s} + \mathcal{D}_{2k+1;2s+1}(r)q_{s}\Big) = -(\epsilon -1 -U(r))u_{k} \ , \\
\end{split}
\end{equation}
while for an odd one
\begin{equation}\label{eq:63od}
\begin{split}
	\partial_r u_{k} &- \frac{2k+1}{r}u_{k} + \lambda \sum_{s} \Big( \mathcal{A}_{2k+1;2s+1}(r)u_{s} + \mathcal{B}_{2k+1;2s+1}(r)v_{s}\Big) = (\epsilon +1 -U(r))q_{k} \\
	\partial_r v_{k} &+ \frac{2k+2}{r}v_{k} + \lambda \sum_{s} \Big( \mathcal{C}_{2k+1;2s+1}(r)u_{s} + \mathcal{D}_{2k+1;2s+1}(r)v_{s}\Big) = (\epsilon +1 -U(r))p_{k} \\	
	\partial_r p_{k} &- \frac{2k}{r}p_{k} - \lambda \sum_{s} \Big( \mathcal{A}_{2k;2s}(r)p_{s} + \mathcal{B}_{2k;2s+2}(r)q_{s}\Big) = -(\epsilon -1 -U(r))v_{k} \\
	\partial_r q_{k} &+ \frac{2k+3}{r}q_{k} - \lambda \sum_{s} \Big( \mathcal{C}_{2k+2;2s}(r)p_{s} + \mathcal{D}_{2k+2;2s+2}(r)q_{s}\Big) = -(\epsilon -1 -U(r))u_{k}\,, \\
\end{split}
\end{equation}
\end{subequations}
where the coefficient  functions $\mathcal{A}(r), \mathcal{B}(r), \mathcal{C}(r), \mathcal{D}(r)$
stand for the matrix elements of commutators $[\, \vec{\sigma} \vec{p} \,, V(\vec{r}\,) ]$ over the spherical spinors
\begin{align}\label{eq:64}
\mathcal{A}_{l;s}(r)  &= i\langle \Omega^{(-)}_{l+1,m_j} | [\, \vec{\sigma} \vec{p} \,, V(\vec{r}\,) ] | \Omega^{(+)}_{s,m_j} \rangle, &
\mathcal{C}_{l;s}(r)  &= i\langle \Omega^{(+)}_{l-1,m_j} | [\, \vec{\sigma} \vec{p} \,, V(\vec{r}\,) ] | \Omega^{(+)}_{s,m_j} \rangle, \nonumber \\
\mathcal{B}_{l;s}(r)  &= i\langle \Omega^{(-)}_{l+1,m_j} | [\, \vec{\sigma} \vec{p} \,, V(\vec{r}\,) ] | \Omega^{(-)}_{s,m_j} \rangle, &
\mathcal{D}_{l;s}(r)  &= i\langle \Omega^{(+)}_{l-1,m_j} | [\, \vec{\sigma} \vec{p} \,, V(\vec{r}\,) ] | \Omega^{(-)}_{s,m_j} \rangle ,
\end{align}
and could be reduced to matrix elements of the form $\langle \Omega^{(\pm)}_{l,m_j} | V(\vec{r}\,) | \Omega^{(\mp)}_{s,m_j} \rangle$. Calculation of the latter is performed by  taking into account the axial symmetry of the potential $V(\vec{r}\,)$, that implies
\begin{equation}\label{eq:65}
V(\vec{r}\,)= \sum_{n} G_{n}(r) P_n(\cos \vartheta) \, ,
\end{equation}
whence for the functions $\mathcal{A}(r), \mathcal{B}(r), \mathcal{C}(r), \mathcal{D}(r)$ one obtains the following expressions
\begin{align}\label{eq:66}
\mathcal{A}_{l;s}(r)  &= \sum_{n=|l-s|}^{|l+s|} \left( \partial_r  - \frac{l-s}{r} \right) G_{n}(r)\,  W^+_{+}(n;l;s),  \nonumber \\
\mathcal{B}_{l;s}(r)  &= \sum_{n=|l-s|}^{|l+s|} \left( \partial_r  -\frac{l+s+1}{r} \right) G_{n}(r)\, W^+_{-}(n;l;s),  \nonumber \\
\mathcal{C}_{l;s}(r)  &= \sum_{n=|l-s|}^{|l+s|} \left( \partial_r  + \frac{l+s+1}{r} \right) G_{n}(r)\, W^-_{+}(n;l;s),  \nonumber \\
\mathcal{D}_{l;s}(r)  &= \sum_{n=|l-s|}^{|l+s|} \left( \partial_r  + \frac{l-s}{r} \right) G_{n}(r)\, W^-_{-}(n;l;s),
\end{align}
where the coefficients $W^\pm_\mp (n;l;s)\equiv \langle \Omega^{(\pm)}_{l,m_j} | P_n(\cos \vartheta) | \Omega^{(\mp)}_{s,m_j} \rangle $  are given through the following combinations of $3j$-symbols
\begin{equation*} \begin{split}
	W^+_\pm(n;l;s)=\sqrt{(l+m_j+\myfrac{1}{2})(s\pm m_j+\myfrac{1}{2})} \ w_n^-(l;s) \pm
	 \sqrt{(l-m_j+\myfrac{1}{2})(s\mp m_j+\myfrac{1}{2})} \ w_n^+(l;s)\\
	W^-_\pm(n;l;s)=\sqrt{(l-m_j+\myfrac{1}{2})(s\pm m_j+\myfrac{1}{2})} \ w_n^-(l;s) \mp  \sqrt{(l+m_j+\myfrac{1}{2})(s\mp m_j+\myfrac{1}{2})} \ w_n^+(l;s)\\
\end{split}
\end{equation*}
\begin{equation}
w_n^\pm(l;s)=(-1)^{m_j\pm \myfrac{1}{2}}  \begin{pmatrix} l & n & s \\ -(m_j\pm \myfrac{1}{2}) & 0 & m_j \pm \myfrac{1}{2} \end{pmatrix} \begin{pmatrix} l & n & s \\ 0 & 0 & 0 \end{pmatrix} \ .
\end{equation}
The coefficient functions $G_n(r)$, entering  the expansion~\eqref{eq:65}, are found via the integral representation
\begin{equation}\label{eq:67}
G_n(r)=\frac{2n+1}{2} \int\limits_{0}^{\pi} \!\sin \theta  \,d\theta \, P_n(\cos \theta) V(r, \tt )\, .
\end{equation}
Since the potential $V(\vec{r}\, )$ is even, $G_n(r)$ don't vanish for even $n$ only. Upon taking account of the explicit form of $c(r)$ for the point-like source~\eqref{eq:10} and the expansion
\begin{equation}\label{eq:68}
\frac{e^{ik|\vec{r}-\vec{a}|}}{|\vec{r}-\vec{a}|} = \frac{i \pi}{2 \sqrt{ra}}\sum_{l=0}^\infty (2l+1) J_{l+1/2}(k r_<) H^{(1)}_{l+1/2} (kr_>) P_l(\cos \theta)\, ,
\end{equation}
where $r_<=\min(r,a)$, $r_>=\max(r,a)$, as well as the orthogonality conditions for the Legendre polynomials, the final answer for the functions~\eqref{eq:67} for even $n$ could be written in the form
\begin{equation}\label{eq:69}
G_n(r)=2Z_q\left(\frac{r_<^n}{r_>^{n+1}}-\frac{2n+1}{2\sqrt{ra}}\int\limits_{4m^2}^\infty \!\frac{dQ^2}{Q^2}\, \frac{\text{Im}\, F_2(Q^2)}{F_2(0)}\, i\,J_{n+1/2}(iQr_<) \,H^{(1)}_{n+1/2}(iQr_>)\right)\, , \quad n=\text{even} \, .
\end{equation}

The calculations, performed for a pair of $u$- or $d$- quarks, show that the level shift  $\d \e_{qq}(a)$ due to displacement from the center is proportional for such pairs to the total charge of quarks, while the curves $|(2 \D \e_q +\d \e_{qq}(a))/2 Z_q|$ coincide almost exactly with those, shown in Fig.~\ref{pic:5}. The configurations, containing a more number of quarks, like $uuu\, , ddd\, , udu\, , dud$ with an additional central quark, don't change this picture. As a result, the total level shift, found quite similarly to~\eqref{eq:55}, amounts to $D \e  = 1.118$~KeV for the level $1s_{1/2}$ with $Z=170$ and $D \e = -1.089$~KeV for the level $2p_{1/2}$ with $Z=183$, which best matches with another non-perturbative result~\eqref{eq:28}, when the whole extended nucleus is treated as the source for $\Delta U_{AMM}$, and also with  results accounting for displacement via PT with unperturbed WF $\psi^{(q)}_{nj}$ including one quark in the center. So the non-perturbative approach to accounting for $\Delta U_{AMM}$ within DE~\eqref{eq:18} reveals a slight advantage over  purely perturbative methods. Thus, the best approximation to the exact result for the shift of the energy level $nj$, taking into account many-quark configurations also, turns out to be the approximation of uniformly charged nucleus as the source of $\D U_{AMM}$~\eqref{eq:28}, as well as the calculation  via  \eqref{eq:31}-\eqref{eq:55}, which coincides almost exactly with the former. And indeed these approximations will be used in further analysis.

\section{General properties of the lowest levels shifts due to $\Delta U_{AMM}$\label{sec:Bc}}

Now let us consider the dependence of energy levels caused by $\Delta U_{AMM}$ on $Z$ near the threshold of the lower continuum for a substantially more wide range  of $Z$, but at the same time in a more qualitatively approach. For an atomic electron the shift, stipulated by $\Delta U_{AMM}$, is a part of the self-energy contribution to the Lamb shift, which in the perturbative QED is proportional to $Z^4/n^3$~\cite{Mohr1998} and is usually represented in terms of the function $F_{nj}(Z\alpha)$, defined by
\beq \label{eq:70}
\D E^{SE}_{nj}(Z\a)=\frac{ Z^4 \a^5 }{ \pi n^3} F_{nj}(Z\a)  \, .
\eeq
In the perturbative QED $F_{nj}(Z\alpha)$ is found for the lowest electronic levels of H-like atoms  with the nucleus charge in the range $Z=1-110$ for all orders in $Z\a$~\cite[and refs. therein]{Mohr1974, Mohr1982, Mohr1992, Johnson1985, Yerokhin2015}. In the case of $Z\alpha > 1$ the calculation of $\D E^{SE}_{1s_{1/2}}$ with the precision of certain percents for the nucleus charge $Z=140,150,160,170$ is given in~\cite{Cheng1976, Soff1982}.

\begin{figure}
\subfigure[]{ \label{pic:4a}
\includegraphics[width=.48\textwidth]{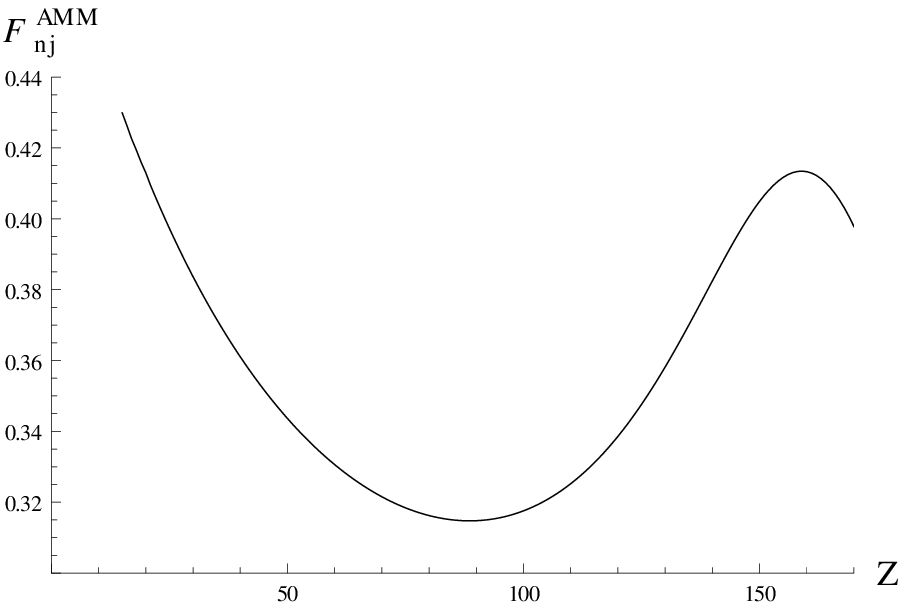}
}
\hfill
\subfigure[]{ \label{pic:4b}
\includegraphics[width=.48\textwidth]{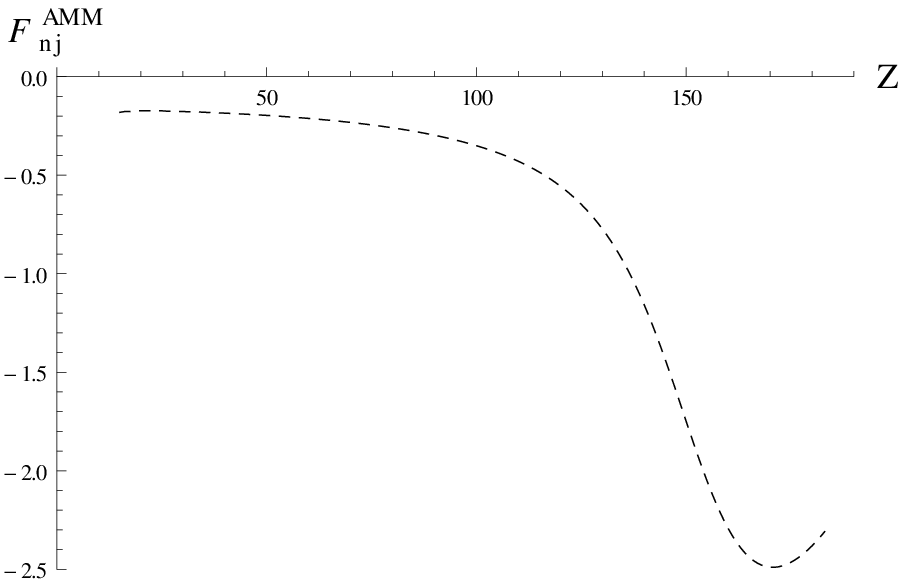}
}
\caption{The function $F_{nj}^{AMM}(Z\a)$ for the shift of levels $1s_{1/2}$ (solid line) and $2p_{1/2}$ (dotted line) due to $\D U_{AMM}$.}\label{pic:4}
\end{figure}

For the Dirac-Pauli operator~\eqref{eq:1} the perturbative calculations of the contribution caused by $\Delta U_{AMM}$ to  $F_{nj}$ are performed in~\cite{Barut1982} for a point-like nucleus with $Z<137$, and in~\cite{Sveshnikov2016} for an extended nucleus with the same dependence $R(Z)$, as in~\eqref{eq:11}. For small $Z$ the behavior of $F_{nj}^{AMM}(Z\a)$ for the Dirac-Pauli operator practically coincides both for the point-like and extended nucleus, while for the increasing nucleus charge $Z$ they grow in a different way, but both quite fast. Moreover,   for a point-like nucleus $F_{nj}^{AMM}(Z\a)$ reveals two poles at subcritical  $Z\a=\sqrt{3}/2$ and critical $Z\a = 1$ values of the nucleus charge~\cite{Barut1982}. However, for an extended nucleus with account for the effective dependence of AMM on the distance~\eqref{eq:16} $F_{nj}^{AMM}(Z\a)$ behaves differently. In Fig.~\ref{pic:4} the function $F_{nj}^{AMM}(Z\a)$ for the levels $1s_{1/2}$ and $2p_{1/2}$ is shown, evaluated within the non-perturbative approach based on~\eqref{eq:28} and~\eqref{eq:55} with practically the same result. In particular, there follows from Fig.~\ref{pic:4}, that  accounting for the dependence of the electronic formfactor on the momentum transfer for the lowest level $1s_{1/2}$ leads to qualitative coincidence (up to numerical factor) between    the behavior of $F_{1s_{1/2}}^{AMM}(Z\a)$ and $F_{1s_{1/2}}(Z\a)$ for the total self-energy shift, namely, it decreases with growing nucleus charge up to $Z\sim 90$, after which starts to increase (compare with~\cite{Cheng1976, Mohr1998}).

%
%

\begin{figure}
\center
\includegraphics[width=.49\textwidth]{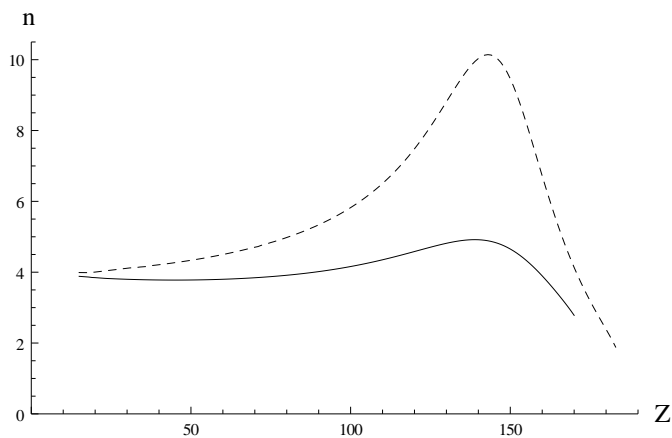}
\caption{The rate of growth $n(Z)$ for the shift of levels $1s_{1/2}$ (solid line) and $2p_{1/2}$ (dotted line) due to $\D U_{AMM}$.}\label{pic:3}
\end{figure}

Another important feature of QED-effects is characterized by the form of their power-low dependence on $Z$.
In Fig.~\ref{pic:3} the behavior of the rate of growth $n(Z)$ for the shifts of levels $1s_{1/2}$ and $2p_{1/2}$ due to $\Delta U_{AMM}$ is shown as a function of $Z$, determined via logarithmic derivative
\beq \label{eq:71}
n(Z)= Z \frac{  \pd }{ \pd Z} \ln \(| \mathcal{D} \e_{AMM}| \) \, .
\eeq

%


There follows from Fig.~\ref{pic:3}, that in this case the behavior of $n(Z)$ in accordance with~\eqref{eq:70}, i.e. the increase of QED-effects $\sim Z^4$,  takes place up to $Z \sim 60-80$, while $n(Z)$ for $1s_{1/2}$ has a shallow minimum at $Z \sim 50$ and tends to $n=4$ for $Z\rightarrow0$ from below. However,  further the shift from  $\D U_{AMM}$ reveals a substantial increase of the growth rate, which reaches its maximum for $Z \simeq 147$ both for $1s_{1/2}$ and $2p_{1/2}$. Moreover, this maximum is much more pronounced for $2p_{1/2}$. Generally speaking, such behavior for the QED-effects is quite natural, since for $Z\a>1$ a substantial growth of non-perturbativity in $Z\a$ should be expected, while the maximum of the rate of growth for $Z \simeq 147$ is a specific feature  of $\D U_{AMM}$. In particular, for a point-like nucleus and unscreened $\D U_{AMM}$  the first level, which approaches the lower continuum indeed at  $Z\simeq 147$,  turns out to be $2p_{1/2}$, with the same (maximal) rate of level diving into  the lower continuum~\cite{Sveshnikov2013}.

\begin{figure}
\subfigure[]{
\includegraphics[width=.48\textwidth]{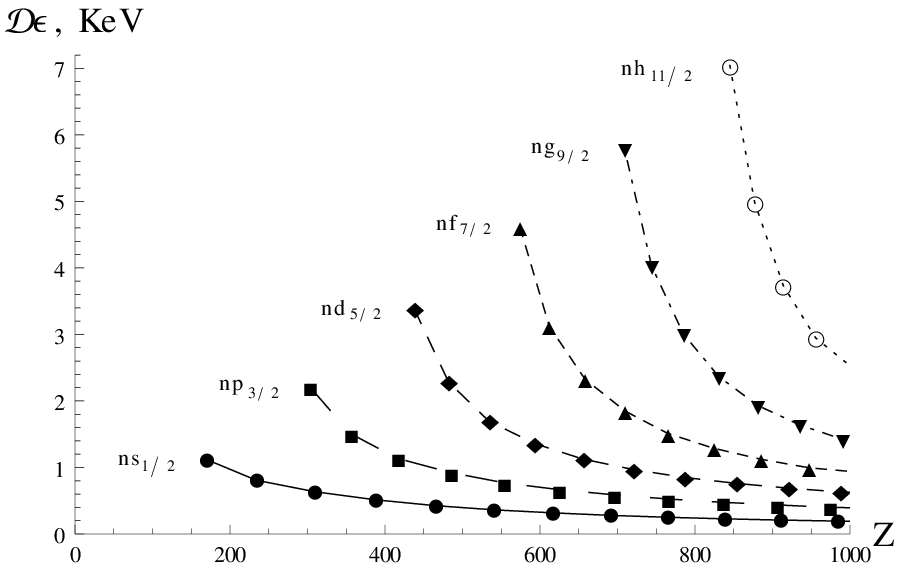} \label{pic:65a}
}
\hfill
\subfigure[]{
\includegraphics[width=.48\textwidth]{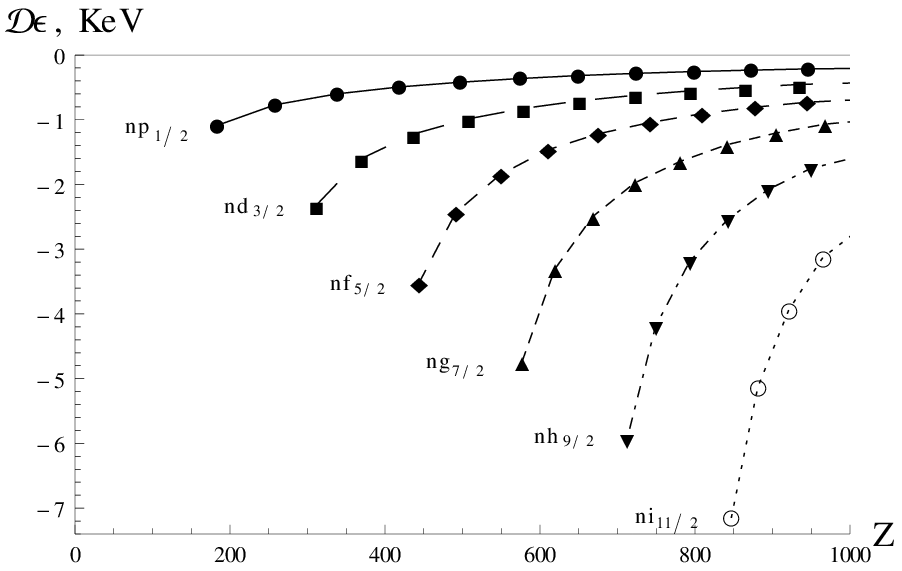} \label{pic:6b}
}
\caption{The shifts of levels with different~$nlj$ due to $\Delta U_{AMM}$, when they approach the threshold of the lower continuum, for the range $Z_{cr,1} < Z < 1000$. The separate trajectories correspond to the levels with fixed parity and $lj$. In Fig.~\subref{pic:65a} the levels with $\k <0\, , \ j=l+\myfrac{1}{2}$ are shown, while in Fig.~\subref{pic:6b} --- with $\k >0\, , \ j=l-\myfrac{1}{2}$ .}\label{pic:6}
\vspace{-.5em}
\end{figure}

Now let us consider the shifts of levels due to $\D U_{AMM}$ for $Z \gg Z_{cr,1}$, when not only the levels  $1s_{1/2}$ and $2p_{1/2}$, but also the levels with another quantum numbers $nlj$ dive into the lower continuum. In particular, for $Z \simeq 234$ the level $2s_{1/2}$ arrives  the threshold of the lower continuum, for $Z \simeq 258$ -- the level $3p_{1/2}$, etc. Direct interest is  the magnitude of radiation effects at the threshold of the lower continuum. In Fig.~\ref{pic:6} there are shown the shifts of Coulomb levels $nlj$ due to $\Delta U_{AMM}$, evaluated via~\eqref{eq:28} and~\eqref{eq:55} for such $Z$ from the range $Z_{cr,1} < Z < 1000$,  when they lye almost at the threshold of the lower continuum with $\e_{nlj} \simeq -1$. In each series of levels with the fixed values of $lj$ and parity the shift of the level $\mathcal{D}\e_{nlj}$ is maximal for the smallest possible $n$, while by increasing $n$ (and so $Z$) $\mathcal{D}\e_{nlj}$ decreases in absolute value. The last effect can be easily understood on the basis of calculations within PT. With growing  $n$ the radial WF $f_{nj}^{(0)}(r)$ and $g_{nj}^{(0)}(r)$ acquire additional zeros, hence, the product $f_{nj}^{(0)}(r)\, g_{nj}^{(0)}(r)$ becomes alternating and begins to oscillate the more strongly the more $n$. And despite the circumstance, that the region, where the radial WF sufficiently differ from zero, enlarges with growing $n$, the resulting values of integrals like~\eqref{eq:29}, which determine the shift caused by $\Delta U_{AMM}$, decrease. Moreover, in each series $nlj$ the greatest  shift of levels at the threshold of the lower continuum doesn't exceed several KeV and grows very slowly  with increasing  $lj$.

However, by itself the decrease of $\mathcal{D}\e_{nlj}$ with growing $n$ isn't quite informative,  the interest is first of all the difference between $\mathcal{D}\e_{nlj}$ and the perturbative result $\sim Z^4/n^3$ for such $Z$. For these purposes in Tab.~\ref{tab:3} there are given the values  $F_{nj}^{AMM}(Z_{cr}\alpha)$ for the shifts of levels due to $\Delta U_{AMM}$ at the threshold of the lower continuum in the range $Z_{cr,1} < Z < 1000$, which decrease monotonically in each series of levels $nlj$ with growing $n$, thence with growing $Z$. The exception is only the lowest levels in the series with $\kappa=-1,-2,-3,-4$ (that means $l=j-\myfrac{1}{2}$ and $j=\myfrac{1}{2},\myfrac{3}{2},\myfrac{5}{2},\myfrac{7}{2}$), for which in the first step  $F_{nj}^{AMM}(Z_{cr}\alpha)$ increase, but in what follows decrease monotonically too. And since it was shown above that with account for the screening of AMM the behavior of $F_{nj}^{AMM}(Z\alpha)$ qualitatively reproduces the behavior of $F_{nj}(Z\alpha)$ for the total shift of the lowest levels, it should be expected, that in the overcritical region the decrease with growing $Z$ could take place for the whole self-energy contribution.
\begin{table}
\center
\caption{$F^{AMM}_{nj}(Z_{cr}\alpha)$ for the shift of levels with different~$nlj$ due to $\Delta U_{AMM}$
at the threshold of the lower continuum in the range $Z_{cr,1} < Z < 1000$.}\label{tab:3}
\begin{ruledtabular}
\begin{tabular}{c *{12}{b{2.3}}}
\mc{$n$ } & \mc { $ns_{1/2}$ } & \mc { $np_{3/2}$ } & \mc { $nd_{5/2}$ } & \mc { $nf_{7/2}$ } & \mc { $ng_{9/2}$ } & \mc { $nh_{11/2}$ } & \mc { $np_{1/2}$ } & \mc { $nd_{3/2}$ } & \mc { $nf_{5/2}$ } & \mc { $ng_{7/2}$ } & \mc { $nh_{9/2}$ } & \mc { $ni_{11/2}$ } \\
\colrule\noalign{\smallskip}
 1 & 0.398 & 0 & 0 & 0 & 0 & 0 & 0 & 0 & 0 & 0 & 0 & 0 \\
 2 & 0.641 & 0.624 & 0 & 0 & 0 & 0 & -2.309 & 0 & 0 & 0 & 0 & 0 \\
 3 & 0.551 & 0.764 & 0.736 & 0 & 0 & 0 & -1.393 & -2.031 & 0 & 0 & 0 & 0 \\
 4 & 0.431 & 0.722 & 0.805 & 0.809 & 0 & 0 & -0.874 & -1.669 & -1.749 & 0 & 0 & 0 \\
 5 & 0.338 & 0.626 & 0.775 & 0.827 & 0.851 & 0 & -0.601 & -1.273 & -1.559 & -1.586 & 0 & 0 \\
 6 & 0.272 & 0.529 & 0.706 & 0.797 & 0.844 & 0.886 & -0.442 & -0.969 & -1.318 & -1.447 & -1.486 & 0 \\
 7 & 0.223 & 0.449 & 0.624 & 0.740 & 0.807 & 0.859 & -0.341 & -0.754 & -1.089 & -1.278 & -1.358 & -1.414 \\
 8 & 0.187 & 0.384 & 0.549 & 0.671 & 0.754 & 0.814 & -0.272 & -0.604 & -0.900 & -1.107 & -1.228 & -1.296 \\
 9 & 0.159 & 0.333 & 0.483 & 0.606 & 0.695 & 0.763 & -0.223 & -0.496 & -0.750 & -0.957 & -1.095 & -1.186 \\
 10 & 0.137 & 0.292 & 0.427 & 0.544 & 0.638 & - & -0.187 & -0.416 & -0.633 & -0.823 & -0.968 & -1.071 \\
 11 & 0.120 & 0.258 & 0.381 & 0.490 & 0.583 & - & -0.160 & -0.355 & -0.543 & -0.714 & -0.855 & -\\
 12 & 0.106 & 0.229 & 0.342 & - & - & - & -0.138 & -0.307 & -0.472 & -0.624 & - & - \\
\end{tabular}
\end{ruledtabular}
\end{table}

\section{Conclusion\label{sec:7}}

Thus, we have shown how with account for the dynamical screening of the electronic AMM at small distances the  electronic levels behave due to  $\D U_{AMM}$ in the superheavy extended nuclei with $Z\a >1$ near the threshold of the lower continuum. The focus was the question, to what extent the results for the shifts of levels within purely perturbative and non-perturbative in  $\a/\pi$ approaches to calculation of corresponding effects within DE~\eqref{eq:18} could be different, provided that in both cases the complete dependence of the electronic WF on $Z\a$ has been taken into account for the very beginning.  It should be stressed once more here, that in this case the non-perturbativity in $\a/\pi$ has nothing to do with the partial summation of the loop expansion for AMM, rather it implies the effects, which cannot be in principle obtained by means of the standard PT.  The interest to this question is stipulated in part by the recent essentially non-perturbative calculations of the vacuum polarization energy for $Z > Z_{cr,1}$~\cite{Davydov2017, Voronina2017}, indicating  an explicitly nonlinear nature of this effect  beyond the scope of PT.

The presented results show that there exists a slight difference between perturbative and non-perturbative approaches to $\D U_{AMM}$ in DE in favor of the latter, slowly growing with increasing $Z$ (see Tabs.~\ref{tab:1},~\ref{tab:2},~\ref{tab:3}). At the same time, the calculation of contribution from $\D U_{AMM}$ via the quark structure and the whole nucleus, concidered as a uniformly charged extended Coulomb source, leads to results, coinciding within accepted precision of calculations.   This confirmes, that in the radiative QED-effects with the virtual photon exchange, including the self-energy contribution, the calculations via standard PT on the basis of uniformly charged nuclei turn out to be a quite good approximation even for superheavy atoms, which is fully consistent with the general conclusion of works~\cite{Indelicato2011, Pyykko2012}.

Moreover, since with account of AMM screening the behavior of $F_{1s_{1/2}}^{AMM}(Z \alpha)$ reproduces qualitatively the behavior of $F_{1s_{1/2}}(Z \alpha)$ for the total self-energy shift  (see Fig.~\ref{pic:4}) in that range of $Z$, where there are reliable results for $F_{1s_{1/2}}(Z \alpha)$, there appears a natural assumption, that in the overcritical region the decrease with growing  $Z$ should take place also for the total self-energy contribution to  the levels shift at the threshold of the lower continuum, and so for the other radiative QED-effects with virtual photon exchange. At the same time, the behavior of the vacuum energy, in which the main role is played by the contribution from fermionic loop, reveals in the overcritical region for $Z$, i.e. beyond the scope of PT,  an essentially nonlinear growth of the effect~\cite{Davydov2017,Voronina2017}, which is quite different from the perturbative behavior and so cannot be compensated by the radiative corrections.


\bibliography{biblio/AMM}

\begin{thebibliography}{27}%
\makeatletter
\providecommand \@ifxundefined [1]{%
 \@ifx{#1\undefined}
}%
\providecommand \@ifnum [1]{%
 \ifnum #1\expandafter \@firstoftwo
 \else \expandafter \@secondoftwo
 \fi
}%
\providecommand \@ifx [1]{%
 \ifx #1\expandafter \@firstoftwo
 \else \expandafter \@secondoftwo
 \fi
}%
\providecommand \natexlab [1]{#1}%
\providecommand \enquote  [1]{``#1''}%
\providecommand \bibnamefont  [1]{#1}%
\providecommand \bibfnamefont [1]{#1}%
\providecommand \citenamefont [1]{#1}%
\providecommand \href@noop [0]{\@secondoftwo}%
\providecommand \href [0]{\begingroup \@sanitize@url \@href}%
\providecommand \@href[1]{\@@startlink{#1}\@@href}%
\providecommand \@@href[1]{\endgroup#1\@@endlink}%
\providecommand \@sanitize@url [0]{\catcode `\\12\catcode `\$12\catcode
  `\&12\catcode `\#12\catcode `\^12\catcode `\_12\catcode `\%12\relax}%
\providecommand \@@startlink[1]{}%
\providecommand \@@endlink[0]{}%
\providecommand \url  [0]{\begingroup\@sanitize@url \@url }%
\providecommand \@url [1]{\endgroup\@href {#1}{\urlprefix }}%
\providecommand \urlprefix  [0]{URL }%
\providecommand \Eprint [0]{\href }%
\providecommand \doibase [0]{http://dx.doi.org/}%
\providecommand \selectlanguage [0]{\@gobble}%
\providecommand \bibinfo  [0]{\@secondoftwo}%
\providecommand \bibfield  [0]{\@secondoftwo}%
\providecommand \translation [1]{[#1]}%
\providecommand \BibitemOpen [0]{}%
\providecommand \bibitemStop [0]{}%
\providecommand \bibitemNoStop [0]{.\EOS\space}%
\providecommand \EOS [0]{\spacefactor3000\relax}%
\providecommand \BibitemShut  [1]{\csname bibitem#1\endcsname}%
\let\auto@bib@innerbib\@empty
\bibitem [{\citenamefont {Barut}\ and\ \citenamefont
  {Kraus}(1975)}]{Barut1975}%
  \BibitemOpen
  \bibfield  {author} {\bibinfo {author} {\bibfnamefont {A.~O.}\ \bibnamefont
  {Barut}}\ and\ \bibinfo {author} {\bibfnamefont {J.}~\bibnamefont {Kraus}},\
  }\href {\doibase 10.1016/0370-2693(75)90696-6} {\bibfield  {journal}
  {\bibinfo  {journal} {Phys. Lett. B}\ }\textbf {\bibinfo {volume} {59}},\
  \bibinfo {pages} {175 } (\bibinfo {year} {1975})}\BibitemShut {NoStop}%
\bibitem [{\citenamefont {Geiger}\ \emph {et~al.}(1988)\citenamefont {Geiger},
  \citenamefont {Reinhardt}, \citenamefont {Muller},\ and\ \citenamefont
  {Greiner}}]{Geiger1988}%
  \BibitemOpen
  \bibfield  {author} {\bibinfo {author} {\bibfnamefont {K.}~\bibnamefont
  {Geiger}}, \bibinfo {author} {\bibfnamefont {J.}~\bibnamefont {Reinhardt}},
  \bibinfo {author} {\bibfnamefont {B.}~\bibnamefont {Muller}}, \ and\ \bibinfo
  {author} {\bibfnamefont {W.}~\bibnamefont {Greiner}},\ }\href {\doibase
  10.1007/BF01294818} {\bibfield  {journal} {\bibinfo  {journal} {Zeitschrift
  fur Physik A Atomic Nuclei}\ }\textbf {\bibinfo {volume} {329}},\ \bibinfo
  {pages} {77} (\bibinfo {year} {1988})}\BibitemShut {NoStop}%
\bibitem [{\citenamefont {Barut}(1990)}]{Barut1990}%
  \BibitemOpen
  \bibfield  {author} {\bibinfo {author} {\bibfnamefont {A.~O.}\ \bibnamefont
  {Barut}},\ }\href {\doibase 10.1007/BF01292863} {\bibfield  {journal}
  {\bibinfo  {journal} {Zeitschrift fur Physik A Atomic Nuclei}\ }\textbf
  {\bibinfo {volume} {336}},\ \bibinfo {pages} {317} (\bibinfo {year}
  {1990})}\BibitemShut {NoStop}%
\bibitem [{\citenamefont {Reitz}\ and\ \citenamefont
  {Mayer}(2000)}]{Reitz2000}%
  \BibitemOpen
  \bibfield  {author} {\bibinfo {author} {\bibfnamefont {J.~R.}\ \bibnamefont
  {Reitz}}\ and\ \bibinfo {author} {\bibfnamefont {F.~J.}\ \bibnamefont
  {Mayer}},\ }\href {\doibase 10.1063/1.533363} {\bibfield  {journal} {\bibinfo
   {journal} {J. Math. Phys.}\ }\textbf {\bibinfo {volume} {41}},\ \bibinfo
  {pages} {4572} (\bibinfo {year} {2000})}\BibitemShut {NoStop}%
\bibitem [{\citenamefont {Greiner}\ \emph {et~al.}(1985)\citenamefont
  {Greiner}, \citenamefont {Mueller},\ and\ \citenamefont
  {Rafelski}}]{Greiner1985a}%
  \BibitemOpen
  \bibfield  {author} {\bibinfo {author} {\bibfnamefont {W.}~\bibnamefont
  {Greiner}}, \bibinfo {author} {\bibfnamefont {B.}~\bibnamefont {Mueller}}, \
  and\ \bibinfo {author} {\bibfnamefont {J.}~\bibnamefont {Rafelski}},\ }\href
  {http://link.springer.com/book/10.1007/978-3-642-82272-8} {\emph {\bibinfo
  {title} {Quantum Electrodynamics of Strong Fields}}},\ \bibinfo {edition}
  {2nd}\ ed.\ (\bibinfo  {publisher} {Springer},\ \bibinfo {address} {Berlin},\
  \bibinfo {year} {1985})\BibitemShut {NoStop}%
\bibitem [{\citenamefont {Ruffini}\ \emph {et~al.}(2010)\citenamefont
  {Ruffini}, \citenamefont {Vereshchagin},\ and\ \citenamefont
  {Xue}}]{Ruffini2010}%
  \BibitemOpen
  \bibfield  {author} {\bibinfo {author} {\bibfnamefont {R.}~\bibnamefont
  {Ruffini}}, \bibinfo {author} {\bibfnamefont {G.}~\bibnamefont
  {Vereshchagin}}, \ and\ \bibinfo {author} {\bibfnamefont {S.-S.}\
  \bibnamefont {Xue}},\ }\href {\doibase 10.1016/j.physrep.2009.10.004}
  {\bibfield  {journal} {\bibinfo  {journal} {Phys. Rept.}\ }\textbf {\bibinfo
  {volume} {487}},\ \bibinfo {pages} {1 } (\bibinfo {year} {2010})},\ \Eprint
  {http://arxiv.org/abs/0910.0974} {arXiv:0910.0974 [astro-ph.HE]} \BibitemShut
  {NoStop}%
\bibitem [{\citenamefont {Rafelski}\ \emph {et~al.}(2016)\citenamefont
  {Rafelski}, \citenamefont {Kirsch}, \citenamefont {Mueller}, \citenamefont
  {Reinhardt},\ and\ \citenamefont {Greiner}}]{Rafelski2016}%
  \BibitemOpen
  \bibfield  {author} {\bibinfo {author} {\bibfnamefont {J.}~\bibnamefont
  {Rafelski}}, \bibinfo {author} {\bibfnamefont {J.}~\bibnamefont {Kirsch}},
  \bibinfo {author} {\bibfnamefont {B.}~\bibnamefont {Mueller}}, \bibinfo
  {author} {\bibfnamefont {J.}~\bibnamefont {Reinhardt}}, \ and\ \bibinfo
  {author} {\bibfnamefont {W.}~\bibnamefont {Greiner}},\ }\href
  {http://arxiv.org/abs/1604.08690} {\bibfield  {journal} {\bibinfo  {journal}
  {ArXiv e-prints}\ } (\bibinfo {year} {2016})},\ \Eprint
  {http://arxiv.org/abs/1604.08690} {arXiv:1604.08690 [hep-ph, nucl-th,
  physics:atom-ph]} \BibitemShut {NoStop}%
\bibitem [{\citenamefont {Schwerdtfeger}\ \emph {et~al.}(2015)\citenamefont
  {Schwerdtfeger}, \citenamefont {Pasteka}, \citenamefont {Punnett},\ and\
  \citenamefont {Bowman}}]{Schwerdtfeger2015}%
  \BibitemOpen
  \bibfield  {author} {\bibinfo {author} {\bibfnamefont {P.}~\bibnamefont
  {Schwerdtfeger}}, \bibinfo {author} {\bibfnamefont {L.~F.}\ \bibnamefont
  {Pasteka}}, \bibinfo {author} {\bibfnamefont {A.}~\bibnamefont {Punnett}}, \
  and\ \bibinfo {author} {\bibfnamefont {P.~O.}\ \bibnamefont {Bowman}},\
  }\href {\doibase 10.1016/j.nuclphysa.2015.02.005} {\bibfield  {journal}
  {\bibinfo  {journal} {Nucl. Phys. A}\ }\textbf {\bibinfo {volume} {944}},\
  \bibinfo {pages} {551 } (\bibinfo {year} {2015})}\BibitemShut {NoStop}%
\bibitem [{\citenamefont {Sveshnikov}\ and\ \citenamefont
  {Khomovskii}(2013)}]{Sveshnikov2013}%
  \BibitemOpen
  \bibfield  {author} {\bibinfo {author} {\bibfnamefont {K.~A.}\ \bibnamefont
  {Sveshnikov}}\ and\ \bibinfo {author} {\bibfnamefont {D.~I.}\ \bibnamefont
  {Khomovskii}},\ }\href {\doibase 10.1134/S1547477113020155} {\bibfield
  {journal} {\bibinfo  {journal} {Phys. Part. Nucl. Lett.}\ }\textbf {\bibinfo
  {volume} {10}},\ \bibinfo {pages} {119} (\bibinfo {year} {2013})}\BibitemShut
  {NoStop}%
\bibitem [{\citenamefont {Davydov}\ \emph {et~al.}(2017)\citenamefont
  {Davydov}, \citenamefont {Sveshnikov},\ and\ \citenamefont
  {Voronina}}]{Davydov2017}%
  \BibitemOpen
  \bibfield  {author} {\bibinfo {author} {\bibfnamefont {A.}~\bibnamefont
  {Davydov}}, \bibinfo {author} {\bibfnamefont {K.}~\bibnamefont {Sveshnikov}},
  \ and\ \bibinfo {author} {\bibfnamefont {Y.}~\bibnamefont {Voronina}},\
  }\href {\doibase 10.1142/S0217751X17500543} {\bibfield  {journal} {\bibinfo
  {journal} {Int. J. Mod. Phys. A}\ }\textbf {\bibinfo {volume} {32}},\
  \bibinfo {pages} {1750054} (\bibinfo {year} {2017})}\BibitemShut {NoStop}%
\bibitem [{\citenamefont {Voronina}\ \emph {et~al.}(2017)\citenamefont
  {Voronina}, \citenamefont {Davydov},\ and\ \citenamefont
  {Sveshnikov}}]{Voronina2017}%
  \BibitemOpen
  \bibfield  {author} {\bibinfo {author} {\bibfnamefont {Y.}~\bibnamefont
  {Voronina}}, \bibinfo {author} {\bibfnamefont {A.}~\bibnamefont {Davydov}}, \
  and\ \bibinfo {author} {\bibfnamefont {K.}~\bibnamefont {Sveshnikov}},\
  }\href@noop {} {\bibfield  {journal} {\bibinfo  {journal} {Phys. Part. Nucl.
  Lett.}\ } (\bibinfo {year} {2017})},\ \bibinfo {note} {in print}\BibitemShut
  {NoStop}%
\bibitem [{\citenamefont {Lautrup}(1976)}]{Lautrup1976}%
  \BibitemOpen
  \bibfield  {author} {\bibinfo {author} {\bibfnamefont {B.}~\bibnamefont
  {Lautrup}},\ }\href {\doibase 10.1016/0370-2693(76)90060-5} {\bibfield
  {journal} {\bibinfo  {journal} {Phys. Lett. B}\ }\textbf {\bibinfo {volume}
  {62}},\ \bibinfo {pages} {103 } (\bibinfo {year} {1976})}\BibitemShut
  {NoStop}%
\bibitem [{\citenamefont {Barut}\ and\ \citenamefont
  {Kraus}(1977)}]{Barut1977}%
  \BibitemOpen
  \bibfield  {author} {\bibinfo {author} {\bibfnamefont {A.~O.}\ \bibnamefont
  {Barut}}\ and\ \bibinfo {author} {\bibfnamefont {J.}~\bibnamefont {Kraus}},\
  }\href {\doibase 10.1103/PhysRevD.16.161} {\bibfield  {journal} {\bibinfo
  {journal} {Phys. Rev. D}\ }\textbf {\bibinfo {volume} {16}},\ \bibinfo
  {pages} {161} (\bibinfo {year} {1977})}\BibitemShut {NoStop}%
\bibitem [{\citenamefont {Itzykson}\ and\ \citenamefont
  {Zuber}(1980)}]{Itzykson1980}%
  \BibitemOpen
  \bibfield  {author} {\bibinfo {author} {\bibfnamefont {C.}~\bibnamefont
  {Itzykson}}\ and\ \bibinfo {author} {\bibfnamefont {J.-B.}\ \bibnamefont
  {Zuber}},\ }\href@noop {} {\emph {\bibinfo {title} {Quantum Field Theory}}}\
  (\bibinfo  {publisher} {McGraw-Hill},\ \bibinfo {year} {1980})\BibitemShut
  {NoStop}%
\bibitem [{\citenamefont {Bateman}\ and\ \citenamefont
  {Erdelyi}(1953)}]{Bateman1953}%
  \BibitemOpen
  \bibfield  {author} {\bibinfo {author} {\bibfnamefont {H.}~\bibnamefont
  {Bateman}}\ and\ \bibinfo {author} {\bibfnamefont {A.}~\bibnamefont
  {Erdelyi}},\ }\href@noop {} {\emph {\bibinfo {title} {Higher Transcendental
  Functions}}},\ Vol.\ \bibinfo {volume} {1-2}\ (\bibinfo  {publisher} {Mc
  Graw-Hill, New York},\ \bibinfo {year} {1953})\BibitemShut {NoStop}%
\bibitem [{\citenamefont {Mohr}\ \emph {et~al.}(1998)\citenamefont {Mohr},
  \citenamefont {Plunien},\ and\ \citenamefont {Soff}}]{Mohr1998}%
  \BibitemOpen
  \bibfield  {author} {\bibinfo {author} {\bibfnamefont {P.~J.}\ \bibnamefont
  {Mohr}}, \bibinfo {author} {\bibfnamefont {G.}~\bibnamefont {Plunien}}, \
  and\ \bibinfo {author} {\bibfnamefont {G.}~\bibnamefont {Soff}},\ }\href
  {\doibase 10.1016/S0370-1573(97)00046-X} {\bibfield  {journal} {\bibinfo
  {journal} {Phys. Rept.}\ }\textbf {\bibinfo {volume} {293}},\ \bibinfo
  {pages} {227 } (\bibinfo {year} {1998})}\BibitemShut {NoStop}%
\bibitem [{\citenamefont {Mohr}(1974)}]{Mohr1974}%
  \BibitemOpen
  \bibfield  {author} {\bibinfo {author} {\bibfnamefont {P.~J.}\ \bibnamefont
  {Mohr}},\ }\href {\doibase 10.1016/0003-4916(74)90398-4} {\bibfield
  {journal} {\bibinfo  {journal} {Ann. Phys.}\ }\textbf {\bibinfo {volume}
  {88}},\ \bibinfo {pages} {26 } (\bibinfo {year} {1974})}\BibitemShut
  {NoStop}%
\bibitem [{\citenamefont {Mohr}(1982)}]{Mohr1982}%
  \BibitemOpen
  \bibfield  {author} {\bibinfo {author} {\bibfnamefont {P.~J.}\ \bibnamefont
  {Mohr}},\ }\href {\doibase 10.1103/PhysRevA.26.2338} {\bibfield  {journal}
  {\bibinfo  {journal} {Phys. Rev. A}\ }\textbf {\bibinfo {volume} {26}},\
  \bibinfo {pages} {2338} (\bibinfo {year} {1982})}\BibitemShut {NoStop}%
\bibitem [{\citenamefont {Mohr}\ and\ \citenamefont {Kim}(1992)}]{Mohr1992}%
  \BibitemOpen
  \bibfield  {author} {\bibinfo {author} {\bibfnamefont {P.~J.}\ \bibnamefont
  {Mohr}}\ and\ \bibinfo {author} {\bibfnamefont {Y.-K.}\ \bibnamefont {Kim}},\
  }\href {\doibase 10.1103/PhysRevA.45.2727} {\bibfield  {journal} {\bibinfo
  {journal} {Phys. Rev. A}\ }\textbf {\bibinfo {volume} {45}},\ \bibinfo
  {pages} {2727} (\bibinfo {year} {1992})}\BibitemShut {NoStop}%
\bibitem [{\citenamefont {Johnson}\ and\ \citenamefont
  {Soff}(1985)}]{Johnson1985}%
  \BibitemOpen
  \bibfield  {author} {\bibinfo {author} {\bibfnamefont {W.~R.}\ \bibnamefont
  {Johnson}}\ and\ \bibinfo {author} {\bibfnamefont {G.}~\bibnamefont {Soff}},\
  }\href {\doibase 10.1016/0092-640X(85)90010-5} {\bibfield  {journal}
  {\bibinfo  {journal} {Atomic Data and Nuclear Data Tables}\ }\textbf
  {\bibinfo {volume} {33}},\ \bibinfo {pages} {405} (\bibinfo {year}
  {1985})}\BibitemShut {NoStop}%
\bibitem [{\citenamefont {Yerokhin}\ and\ \citenamefont
  {Shabaev}(2015)}]{Yerokhin2015}%
  \BibitemOpen
  \bibfield  {author} {\bibinfo {author} {\bibfnamefont {V.~A.}\ \bibnamefont
  {Yerokhin}}\ and\ \bibinfo {author} {\bibfnamefont {V.~M.}\ \bibnamefont
  {Shabaev}},\ }\href {\doibase 10.1063/1.4927487} {\bibfield  {journal}
  {\bibinfo  {journal} {J. Phys. Chem. Ref. Data}\ }\textbf {\bibinfo {volume}
  {44}},\ \bibinfo {pages} {033103} (\bibinfo {year} {2015})}\BibitemShut
  {NoStop}%
\bibitem [{\citenamefont {Cheng}\ and\ \citenamefont
  {Johnson}(1976)}]{Cheng1976}%
  \BibitemOpen
  \bibfield  {author} {\bibinfo {author} {\bibfnamefont {K.~T.}\ \bibnamefont
  {Cheng}}\ and\ \bibinfo {author} {\bibfnamefont {W.~R.}\ \bibnamefont
  {Johnson}},\ }\href {\doibase 10.1103/PhysRevA.14.1943} {\bibfield  {journal}
  {\bibinfo  {journal} {Phys. Rev. A}\ }\textbf {\bibinfo {volume} {14}},\
  \bibinfo {pages} {1943} (\bibinfo {year} {1976})}\BibitemShut {NoStop}%
\bibitem [{\citenamefont {Soff}\ \emph {et~al.}(1982)\citenamefont {Soff},
  \citenamefont {Schl\"uter}, \citenamefont {M\"uller},\ and\ \citenamefont
  {Greiner}}]{Soff1982}%
  \BibitemOpen
  \bibfield  {author} {\bibinfo {author} {\bibfnamefont {G.}~\bibnamefont
  {Soff}}, \bibinfo {author} {\bibfnamefont {P.}~\bibnamefont {Schl\"uter}},
  \bibinfo {author} {\bibfnamefont {B.}~\bibnamefont {M\"uller}}, \ and\
  \bibinfo {author} {\bibfnamefont {W.}~\bibnamefont {Greiner}},\ }\href
  {\doibase 10.1103/PhysRevLett.48.1465} {\bibfield  {journal} {\bibinfo
  {journal} {Phys. Rev. Lett.}\ }\textbf {\bibinfo {volume} {48}},\ \bibinfo
  {pages} {1465} (\bibinfo {year} {1982})}\BibitemShut {NoStop}%
\bibitem [{\citenamefont {Barut}\ and\ \citenamefont
  {Kraus}(1982)}]{Barut1982}%
  \BibitemOpen
  \bibfield  {author} {\bibinfo {author} {\bibfnamefont {A.~O.}\ \bibnamefont
  {Barut}}\ and\ \bibinfo {author} {\bibfnamefont {J.}~\bibnamefont {Kraus}},\
  }\href {\doibase 10.1088/0031-8949/25/4/010} {\bibfield  {journal} {\bibinfo
  {journal} {Phys. Scr.}\ }\textbf {\bibinfo {volume} {25}},\ \bibinfo {pages}
  {561} (\bibinfo {year} {1982})}\BibitemShut {NoStop}%
\bibitem [{\citenamefont {Sveshnikov}\ and\ \citenamefont
  {Khomovsky}(2016)}]{Sveshnikov2016}%
  \BibitemOpen
  \bibfield  {author} {\bibinfo {author} {\bibfnamefont {K.~A.}\ \bibnamefont
  {Sveshnikov}}\ and\ \bibinfo {author} {\bibfnamefont {D.~I.}\ \bibnamefont
  {Khomovsky}},\ }\href {\doibase 10.3103/S0027134916050179} {\bibfield
  {journal} {\bibinfo  {journal} {Moscow Uni Phys. Bull.}\ }\textbf {\bibinfo
  {volume} {71}},\ \bibinfo {pages} {3} (\bibinfo {year} {2016})}\BibitemShut
  {NoStop}%
\bibitem [{\citenamefont {Indelicato}\ \emph {et~al.}(2011)\citenamefont
  {Indelicato}, \citenamefont {Biero{\'{n}}},\ and\ \citenamefont
  {J{\"o}nsson}}]{Indelicato2011}%
  \BibitemOpen
  \bibfield  {author} {\bibinfo {author} {\bibfnamefont {P.}~\bibnamefont
  {Indelicato}}, \bibinfo {author} {\bibfnamefont {J.}~\bibnamefont
  {Biero{\'{n}}}}, \ and\ \bibinfo {author} {\bibfnamefont {P.}~\bibnamefont
  {J{\"o}nsson}},\ }\href {\doibase 10.1007/s00214-010-0887-3} {\bibfield
  {journal} {\bibinfo  {journal} {Theor. Chem. Acc.}\ }\textbf {\bibinfo
  {volume} {129}},\ \bibinfo {pages} {495} (\bibinfo {year}
  {2011})}\BibitemShut {NoStop}%
\bibitem [{\citenamefont {Pyykk{\"o}}(2012)}]{Pyykko2012}%
  \BibitemOpen
  \bibfield  {author} {\bibinfo {author} {\bibfnamefont {P.}~\bibnamefont
  {Pyykk{\"o}}},\ }\href {\doibase 10.1021/cr200042e} {\bibfield  {journal}
  {\bibinfo  {journal} {Chem. Rev.}\ }\textbf {\bibinfo {volume} {112}},\
  \bibinfo {pages} {371–384} (\bibinfo {year} {2012})}\BibitemShut {NoStop}%
\end{thebibliography}%

\end{document}